\documentclass[sigconf,balance=false]{acmart}
\usepackage{popets}
\usepackage{enumitem}

\usepackage{graphicx}
\usepackage{textcomp}
\usepackage{xcolor}

\usepackage{url,caption,subcaption,graphicx}
\usepackage{tabularx,soul}

\usepackage{array}
\newcolumntype{P}[1]{>{\centering\arraybackslash}p{#1}}
\newcolumntype{M}[1]{>{\centering\arraybackslash}m{#1}}

\usepackage{textcomp}
\usepackage{pgfplots}
\usepackage{tcolorbox}

\pgfplotsset{width=8.4cm,height=6cm,compat=1.17}
\usepackage{amsmath,xcolor,soul, fontawesome,algpseudocode,algorithm,xspace,multirow,ctable,longtable,colortbl,lscape,pifont}

\sethlcolor{red}
\usepackage[utf8]{inputenc}
\usepackage[font={small}]{caption}
\usepackage{listings}
\usepackage{xcolor}
\usepackage{pifont}
\usepackage{tikz}

\definecolor{codegreen}{rgb}{0,0.6,0}
\definecolor{codegray}{rgb}{0.5,0.5,0.5}
\definecolor{codepurple}{rgb}{0.58,0,0.82}
\definecolor{backcolour}{rgb}{0.95,0.95,0.92}
\definecolor{bblue}{HTML}{4F81BD}
\definecolor{rred}{HTML}{E11916}
\definecolor{ggreen}{HTML}{3FD72D}
\definecolor{ggreen1}{HTML}{9DEC9D}
\definecolor{ppurple}{HTML}{9F4C7C}
\definecolor{yyellow}{HTML}{FFC000}
\definecolor{yyellow1}{HTML}{FEE599}

\definecolor{debug}{HTML}{FFBABA}
\definecolor{info}{HTML}{FF5252}
\definecolor{warning}{HTML}{FF0000}
\definecolor{severe}{HTML}{A70000}

\definecolor{last-year}{HTML}{1E476C}
\definecolor{this-year}{HTML}{9FC5E8}

\lstdefinestyle{mystyle}{
  backgroundcolor=\color{backcolour},   commentstyle=\color{codegreen},
  keywordstyle=\color{magenta},
  numberstyle=\tiny\color{codegray},
  stringstyle=\color{codepurple},
  basicstyle=\ttfamily\footnotesize,
  breakatwhitespace=false,         
  breaklines=true,                 
  captionpos=b,                    
  keepspaces=true,                 
  numbers=left,                    
  numbersep=5pt,                  
  showspaces=false,                
  showstringspaces=false,
  showtabs=false,                  
  tabsize=1 
}

\usepackage{booktabs,siunitx,comment}

\newcommand{\ie}{\emph{i.e.,}\xspace}
\newcommand{\eg}{\emph{e.g.,}\xspace}

\setcopyright{popets}
\copyrightyear{YYYY}

\acmYear{YYYY}
\acmVolume{YYYY}
\acmNumber{X}
\acmDOI{XXXXXXX.XXXXXXX}
\acmISBN{}
\acmConference{Proceedings on Privacy Enhancing Technologies}
\begin{document}

\title[Blocking JavaScript Without Breaking the Web: An Empirical Investigation]{Blocking JavaScript Without Breaking the Web: \\An Empirical Investigation}

\author{Abdul Haddi Amjad}
\affiliation{%
  \institution{Virginia Tech}
  \city{}
  \state{}
  \country{}
  }
\email{hadiamjad@vt.edu}

\author{Zubair Shafiq}
\affiliation{%
  \institution{University of California, Davis}
  \city{}
  \state{}
  \country{}
  }
\email{zubair@ucdavis.edu}

\author{Muhammad Ali Gulzar}
\affiliation{%
  \institution{Virginia Tech}
  \city{}
  \state{}
  \country{}
}
\email{gulzar@cs.vt.edu}

\begin{abstract}
Modern websites heavily rely on JavaScript (JS) to implement legitimate functionality as well as privacy-invasive advertising and tracking.
Browser extensions such as NoScript block any script not loaded by a trusted list of endpoints, thus hoping to block privacy-invasive scripts while avoiding breaking legitimate website functionality. 
In this paper, we investigate whether blocking JS on the web is feasible without breaking legitimate functionality. 
To this end, we conduct a large-scale measurement study of JS blocking on 100K websites. 
We evaluate the effectiveness of different JS blocking strategies in tracking prevention and functionality breakage. 
Our evaluation relies on quantitative analysis of network requests and resource loads as well as manual qualitative analysis of visual breakage. 
First, we show that while blocking all scripts is quite effective at reducing tracking, it significantly degrades functionality on approximately two-thirds of the tested websites. 
Second, we show that selective blocking of a subset of scripts based on a curated list achieves a better trade-off. However, there remain approximately 15\% ``mixed'' scripts, which essentially merge tracking and legitimate functionality and thus cannot be blocked without causing website breakage.
Finally, we show that fine-grained blocking of a subset of JS methods, instead of scripts, reduces major breakage by 3.8$\times$ while providing the same level of tracking prevention. 
Our work highlights the promise and open challenges in fine-grained JS blocking for tracking prevention without breaking the web. 
\end{abstract}

\keywords{privacy, web, software engineering}

\maketitle

\begin{figure*}[!t]
\begin{subfigure}{.19\textwidth}
    \includegraphics[width=.99\linewidth]{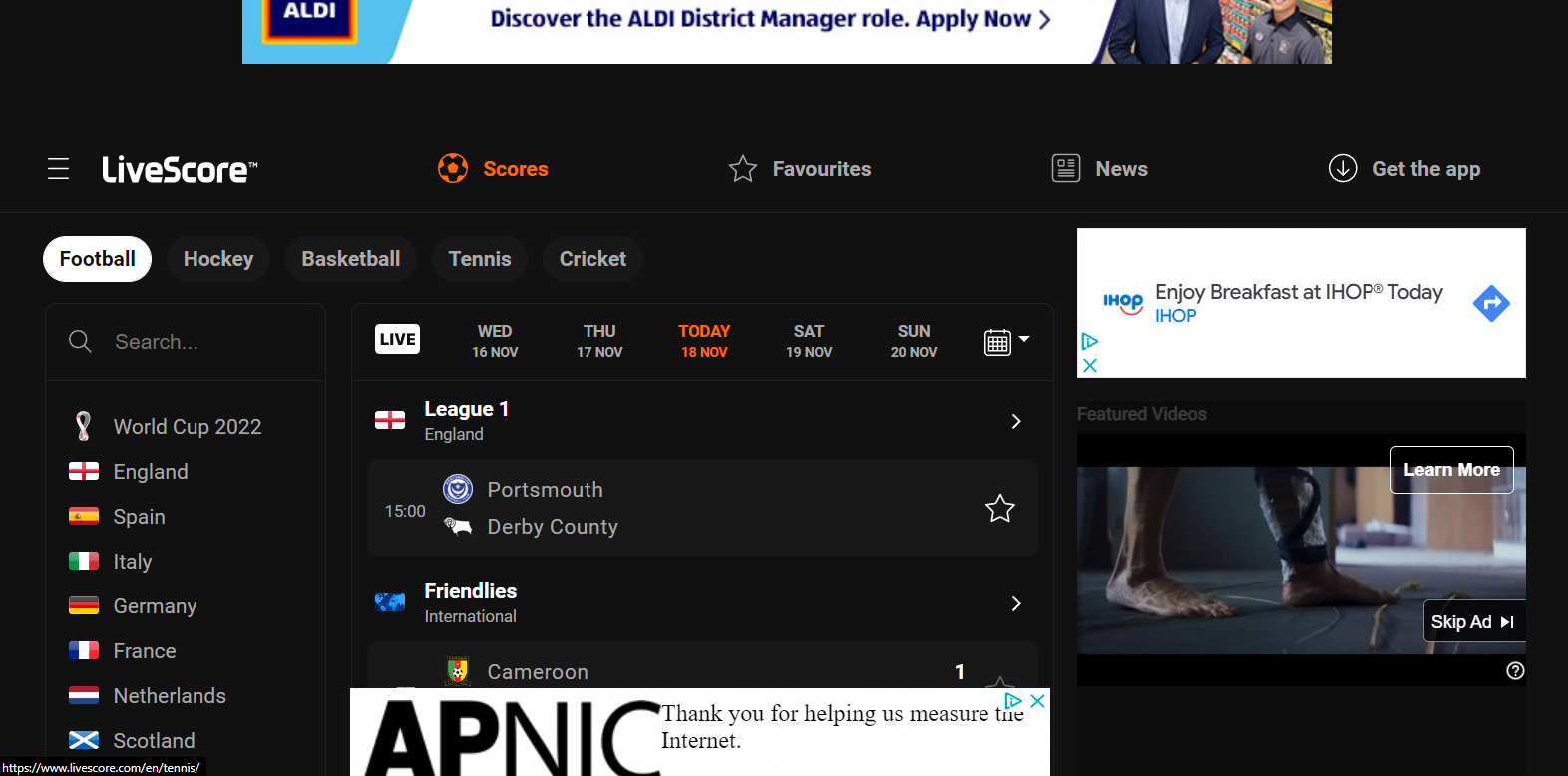}
    \caption{Control}
    \label{fig:livescore-control}
\end{subfigure}
\begin{subfigure}{.19\textwidth}
    \includegraphics[width=.99\linewidth]{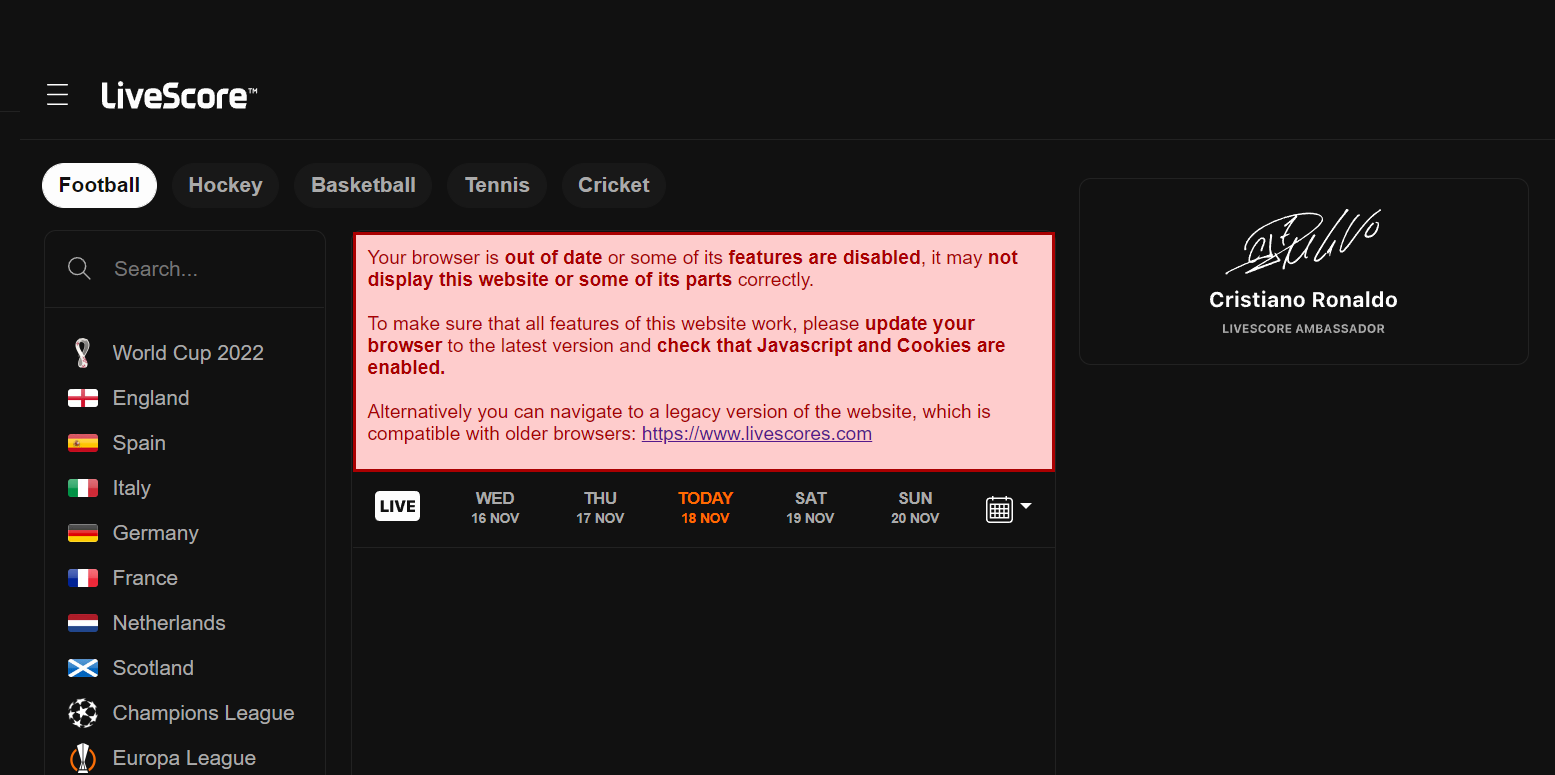}
    \caption{NoScript}
\end{subfigure}
\begin{subfigure}{.19\textwidth}
    \includegraphics[width=.99\linewidth]{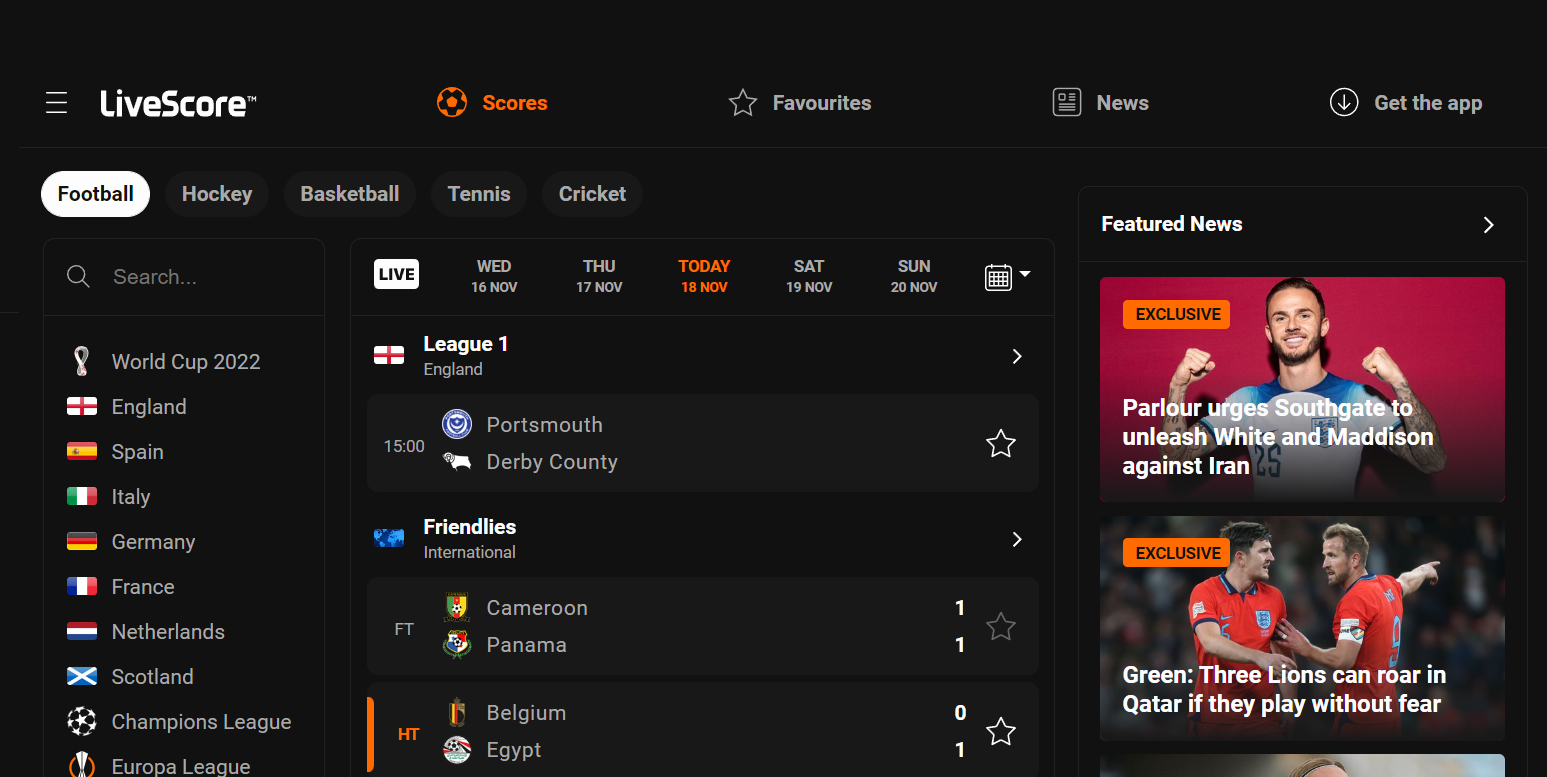}
    \caption{uBlock Origin}
\end{subfigure}
\begin{subfigure}{.19\textwidth}
    \includegraphics[width=.99\linewidth]{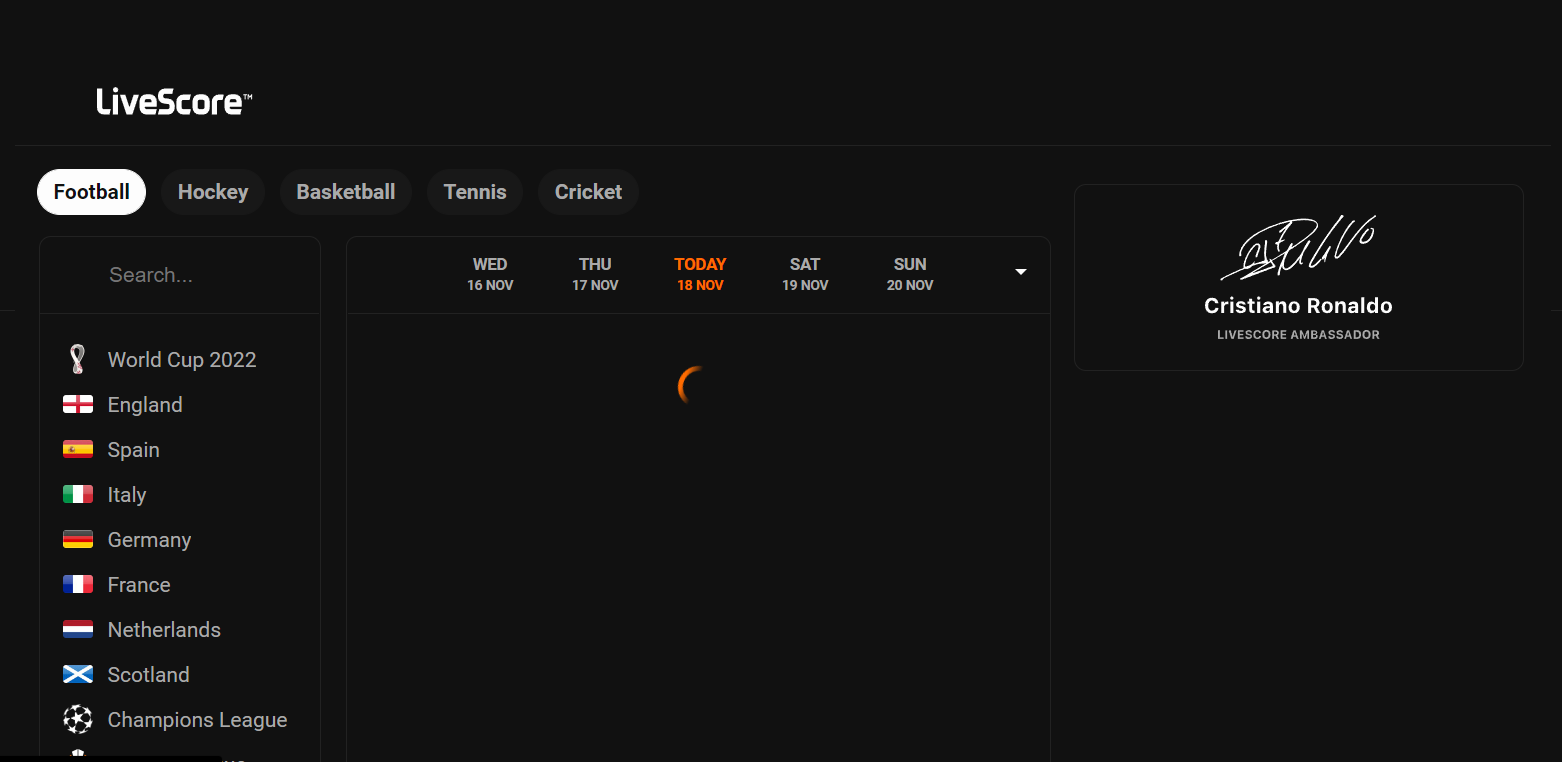}
    \subcaption{Mixed}
    \end{subfigure}
\begin{subfigure}{.19\textwidth}
    \includegraphics[width=.99\linewidth]{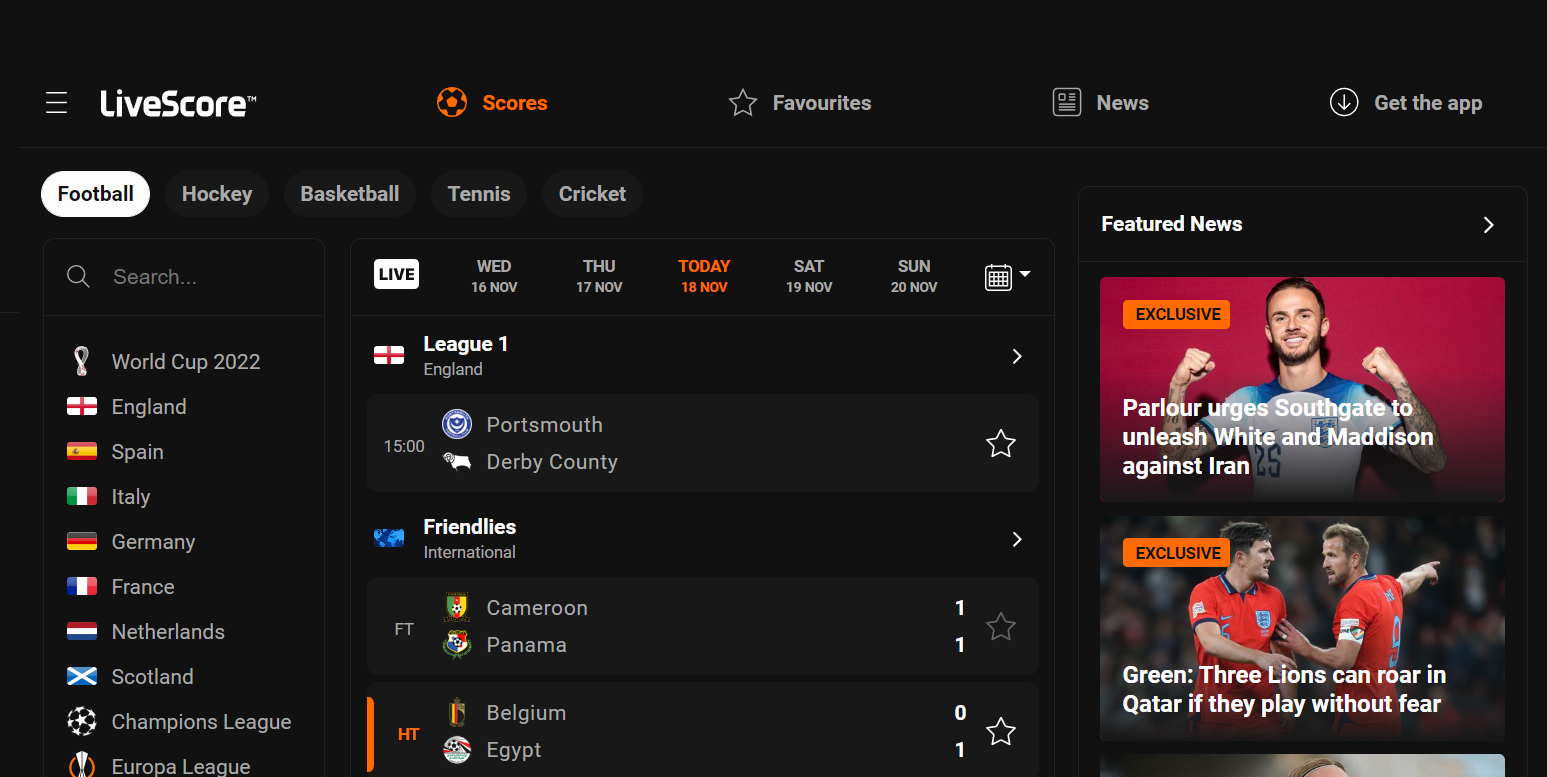}
    \caption{Method}
    \end{subfigure}

\caption{The snapshot of {\tt livescore.com} with (a) control-setting (no content blocker), (b) NoScript (default setting), (c) uBlock Origin (default setting), (d) mixed script blocked ({\tt \_app-*.js}), and (e) JS method blocked (method in {\tt \_app-*.js}).} 
\vspace{-3ex}
\label{fig:livescore}
\end{figure*} 

\section{Introduction}
\label{sec: introduction}
JavaScript is often used to provide rich user experiences on the web. 
The volume of JavaScript on the web has steadily increased over the years. 
The median web page load today ships 500+ kilobytes of JavaScript \cite{goel_2022}. 
While some of it is used to implement various libraries and frameworks (\eg jQuery, React), almost half of it is third-party scripts that implement advertising and tracking services.
The research community is concerned about the negative impact of JavaScript on performance \cite{vazquez2019slimming,chaqfeh2020jscleaner,kupoluyi2021muzeel}, security \cite{mao2018detecting,curtsinger2011zozzle,xu2013jstill,fass2019jstap}, and privacy \cite{jueckstock2019visiblev8,Merzdovnik17BlockMeIfYouCanESP,iqbal21fpinspector,mayer2012third,englehardt2016online}.

Due to these concerns, there is a small but active community of web users who want to use the web without JavaScript. 
In fact, all major browsers now provide a native way for users to block all JavaScript \cite{browserdisablejs}. 
Moreover, users can employ browser extensions such as NoScript \cite{NoScript} that block all scripts -- except those from a trusted source.
HTML5 now also supports the \texttt{noscript} element that allows web developers to gracefully support such browsers that do not support scripting \cite{noscriptelement}.

While blanket JavaScript blocking does alleviate these concerns, it inevitably breaks the legitimate website functionality. 
The privacy community has developed content-blocking tools that selectively block tracking resources (\eg scripts) on a webpage.
Privacy-enhancing content blockers, such as uBlock Origin \cite{ublockOrigin}, block network requests to known trackers by matching request URLs with manually curated filter lists \cite{EasyList1,EasyPrivacy}.

Since these privacy-enhancing content blockers are now used by more than one-third of web users \cite{backlinko_2021, 10.1145/2987443.2987460}, there are strong financial incentives for web developers to evade content blockers. 
The typical evasion strategy is to manipulate the URLs, \eg change the URL path or hostname such that filter lists are no longer effective \cite{Alrizah19IMCerrorsMisunderstandings,iqbal17adwars}. 
This has led to an arms race where filter lists must be promptly updated in response to such evasion attempts \cite{le2022autofr,sugarcoat,chen21jssignatures}. 
Filter list curators have also made a concerted effort to selectively block the underlying scripts from downloading or execution that are responsible for initiating tracking requests. 
In response, a new evasion strategy has emerged where web developers attempt to mix tracking and functional code in the same script (\eg JS bundling ~\cite{chen21jssignatures}). 
Privacy-enhancing content blockers risk breaking a webpage if they block such scripts or compromise user privacy if they do not.

Privacy-enhancing content blockers aim to eliminate tracking while preserving website functionality. 
However, if they are forced to choose --- \eg when tracking and functional code is mixed ---they always prioritize functionality preservation. 
This is because most users tend to disable privacy-enhancing content blockers if they break legitimate website functionality. 
Recent research \cite{trackersift,sugarcoat} has shown that many websites now mix functional and tracking code that renders privacy-enhancing content blocking useless.


In this paper, we conduct a first-of-its-kind empirical investigation of JS blocking. 
To this end, we quantitatively and qualitatively evaluate the impact of different granularities of JS blocking on 100K websites. 
Our goal is to assess whether it is feasible to eliminate tracking effectively while preserving website functionality at different granularities of JS code \ie script and method. 
Beyond blanket JS blocking, we first investigate selective blocking of tracking scripts as well as mixed scripts. 
We further expand our investigation to the effectiveness of method-level blocking.

Our large-scale automated analysis of 100K websites reaffirms that blanket JS blocking indeed eliminates tracking, but it also breaks website functionality on approximately two-thirds of the tested websites. 
We then show that selective blocking of tracking scripts mitigates tracking without degrading website functionality, but there remains a significant fraction of scripts that mix tracking and functional behavior. 
Specifically, we find that 14.6\% of the scripts exhibit both tracking and functional (\ie mixed) behavior. 
We then adapt Spectra-based fault localization (SBFL), a popular faulty code localization technique,  to further localize tracking to the constituent methods of these mixed scripts. 
We find that method-level blocking of tracking methods significantly reduces website breakage while providing the same level of tracking prevention.

We also qualitatively analyze a sample of 383 websites under different JS blocking configurations for functionality breakage. 
%
%
We characterize functionality into four components \eg navigation, single sign-on, appearance, and additional functionality, and quantify breakage on 3-levels (none, minor, and major). 
Our evaluation shows that method-level JS blocking is far better at preserving functionality while achieving a similar level of tracking prevention. 
Specifically, we find that script-level JS blocking results in 3.8$\times$ major breakage and 1.5$\times$ minor breakage as compared to method-level JS blocking.


We summarize our key findings and contributions below:
\begin{itemize}

\item We find that method-level JS blocking is able to prevent tracking on par with script-level JS blocking while improving functionality preservation by 3.8$\times$ major breakage and 1.5$\times$ minor breakage.

\item By comparing two web crawls conducted one year apart, we find a 14\% increase in the number of websites that employ mixed scripts on 100K websites.

\item Even at the method-level granularity, there remain 6\% mixed methods that combine tracking and functionality and require even deeper program analysis for effective blocking without breaking functionality. 

\item The data set crawled for this study offers a full-scale view of JS code integration on today's websites, presenting a detailed lineage of tracking, functional, and mixed JS code units across 100K websites.


\end{itemize}

\noindent {\bf Data Availability:} Our  source code and data is available at \url{https://zenodo.org/record/6526537}.

\section{Motivation}
\label{sec:motivation}
In this section, we present a case study to illustrate the tradeoff between tracking prevention and functionality breakage.

\vspace{.05in} \noindent {\bf No JS blocking.}
Let's take the example of {\tt livescore.com}, a top-10 ranked sports website \cite{livescore}.
We first load the homepage of {\tt livesc\\ore.com} in a stock Chrome browser without any JavaScript intervention. 
Loading this webpage results in 294 network requests in 11 seconds, including 83 requests to fetch scripts and 175 requests initiated by these scripts.
For motivation, consider two of these scripts that initiate network requests to \textit{known}\footnote{See, for example, Disconnect tracking protection list \cite{disconnectme}} tracking endpoints: {\tt gtm.js} served by {\tt googletagmanager.com} and {\tt \_app-*.js } served by {\tt livescore.com}. 
{\tt gtm.js} sends network requests to {\tt googleadse\\rvices.com} and {\tt google-analytics.com}.
%
{\tt \_app-*.js} sends network requests to {\tt doubleclick.net}.  
%
Upon careful inspection, we find that {\tt \_app-*.js} also sends a network request to {\tt livescore.com\\/api/announcements/} that includes known tracking cookies such as {\tt \_gads} \cite{chen2021cookie,munir2022cookiegraph}.
While both scripts are responsible for network requests to tracking endpoints, {\tt \_app-*.js} is a mixed script that seems to implement both legitimate website functionality (\eg add media, populate game statistics) and tracking. 
Figure \ref{fig:livescore} (a) shows the homepage of {\tt livescore.com} in the control configuration (without any blocking).

\begin{figure*}[t]
    \centering
      \includegraphics[width=\textwidth]{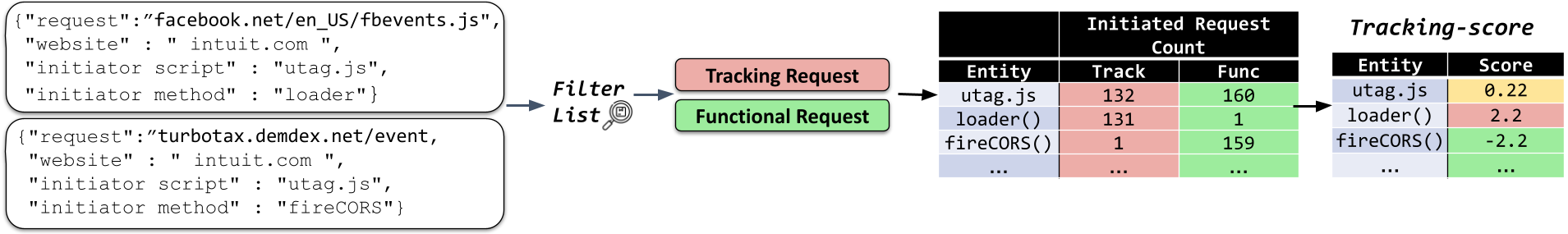}
      \put(-510,75){{\ding{202}}}
      \put(-325,53){{\ding{203}}}
      \put(-150,75){{\ding{204}}}
      \put(-43,72){{\ding{205}}}

        \caption{Steps for localizing tracking and functional JS code using Spectra-based fault localization. \ding{202} shows the two network requests on {\tt intuit.com}. Filter lists are used to label requests in \ding{203}. Spectra-based fault localization is used to classify resources based on participation, as shown in \ding{204} and \ding{205}.}
        \label{fig:annotation}

\end{figure*}

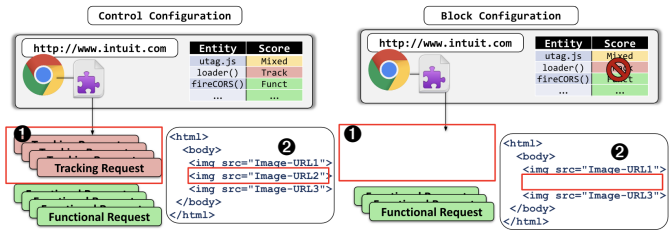
\begin{figure}[t]
    \centering
      \includegraphics[width=0.5\textwidth]{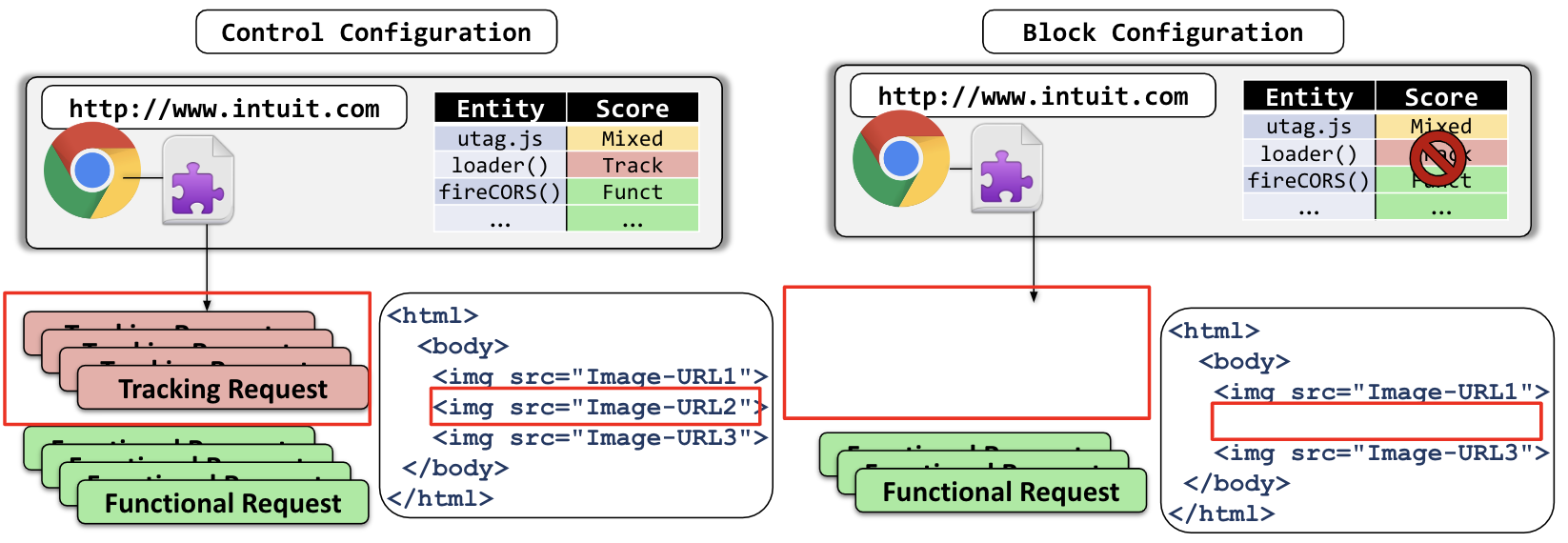}
       \put(-250,35){{\ding{202}}}
       \put(-125,35){{\ding{202}}}
      \put(-150,30){{\ding{203}}}
      \put(-24,28){{\ding{203}}}
        \caption{Illustration of the breakage metrics for automated JS blocking. Request count (\ding{202}) and HTML of website (\ding{203}) are compared with control configuration.}
        \vspace{-5ex}
        \label{fig:metrics}
\end{figure}
\vspace{.05in} \noindent {\bf Blanket JS blocking.}
The naive way is to block all JS on {\tt livescore\\.com} at the page load time.
This capability is available in all major browsers \cite{browserdisablejs}.
While this approach blocks all the aforementioned tracking requests, it also completely breaks the website functionality.
{\tt livescore.com} becomes unusable and in fact notifies the user\footnote{The notice on livescore.com states: ``Your browser is out of date or some of its features are disabled, it may not display this website or some of its parts correctly. To make sure that all features of this website work, please update your browser to the latest version and check that Javascript and Cookies are enabled.''} that JS needs to be enabled for the website to display correctly.
NoScript \cite{NoScript} also blocks all JS on {\tt livescore.com}, including {\tt gtm.js} served by {\tt googletagmanager.com} and {\tt \_app-*.js } served by {\tt livescore.com}.
This again completely breaks the website functionality. 
Figure \ref{fig:livescore} (b) shows the homepage of {\tt livescore.com} when NoScript \cite{NoScript} is used.

\vspace{.05in} \noindent {\bf Selective JS blocking.}
We next use a tracker blocking tool, called uBlock Origin \cite{ublockOrigin}, on {\tt livescore.com}. 
Note that these tracker blocking tools do not specifically target JS. 
Instead, they use a curated filter list to block network requests to known tracking endpoints that  may incidentally include network requests to fetch JS.
Thus, compared to blanket JS blocking, uBlock Origin aims to block all network requests to known tracking endpoints while allowing other network requests. 
After loading {\tt livescore.com} with uBlock Origin installed, we observe that {\tt gtm.js} is blocked, thus eliminating all subsequent tracking network requests from {\tt gtm.js}. 
However, instead of blocking {\tt \_app-*.js}, uBlock Origin blocks the network request to {\tt doubleclick.net} while it allows the  network request {\tt livescore.com/api/announcements/} containing tracking cookies.
Figure \ref{fig:livescore} (c) shows the homepage of {\tt livescore.com} when uBlock Origin \cite{ublockOrigin} is used.
Although there is no website breakage, uBlock Origin has essentially decided not to block {\tt \_app-*.js} to avoid website breakage even though it results in tracking requests. 
As we elaborate later, trackers have been increasingly putting tracker blocking tools in such a bind.


\vspace{.05in} \noindent {\bf Tracking and Mixed JS blocking.}
To understand why uBlock Origin chose not to block {\tt \_app-*.js}, we next use uBlock Origin but also configure it to block {\tt \_app-*.js}. 
As shown in Figure \ref{fig:livescore} (d), this leads to a major functionality breakage on {\tt livescore.com}; the navigation button, game statistics, and the featured news section are not rendered correctly.
Put simply, there is a no-win situation when it comes to {\tt \_app-*.js}. Blocking it results in website breakage, and not blocking it results in tracking.

\lstdefinestyle{base}{
language=Java,
moredelim=**[is][\color{red}]{@}{@},
moredelim=**[is][\color{codegreen}]{!}{!}
}
\begin{lstlisting}[caption={JS method {\tt u} that initiates tracking requests in script {\tt \_app-*.js}. We replace this method name with {\tt donotExecuteMe}.},language=Java, label={lst:mthd_ex}, style=base]
@- u = function(e) {@
!+ donotExecuteMe = function(e) {!
            ...
            return fetch(e).then(c.cg).then((function(e) 
            {return e || {}}))
\end{lstlisting}

\vspace{.05in} \noindent {\bf Method-level JS blocking.} 
Recent work \cite{trackersift, sugarcoat} has applied dynamic analysis to identify tracking methods in mixed scripts manually. 
Our analysis of network requests initiated by {\tt \_app-*.js} shows that the tracking requests were initiated by the method shown in Listing \ref{lst:mthd_ex}. 
As shown in Figure \cite{livescore} (e), when this method in {\tt \_app-*.js} is blocked (\eg it is renamed such that all calls to this method are invalidated), the entire webpage renders completely while all tracking requests are also blocked. 
It is noteworthy that manually refactoring mixed scripts is not feasible at scale. 
Therefore, only a handful of mixed scripts have been refactored in prior work \cite{repl}. 
%

\section{Methodology}
\label{sec:methodology}
This section describes our methodology for automated analysis of JS blocking on 100K webpages (Phase I) and manual inspection of JS blocking on 383 websites (Phase II).

\subsection{Phase I: Automated JS Blocking Analysis}
Figure \ref{fig:annotation} shows our automated JS blocking analysis pipeline comprising a JS collection step and JS code localization step. Figure \ref{fig:metrics} shows our JS blocking impact analysis step.

\subsubsection{JavaScript Corpus Collection}
We crawl landing pages of 100K randomly sampled websites from Tranco top-million list~\cite{pochat2018tranco} using a custom-built Chrome extension.
We spend 20 seconds on a page, exceeding the median {\tt onLoad} time by 13.5 seconds on average. 
This allows us to capture the vast majority of the content fetched, which is consistent with over 90\% of all webpages \cite{onload}. Nonetheless, we measure the impact of increasing the crawl time to 90 seconds on 200 web pages randomly sampled from 100K. We notice average differences of 2\% and 5.2\% in tracking and functional requests, respectively, causing an insignificant impact on our findings. Thus, we set the crawl time to 20 seconds.
%
 
%
For each webpage, our crawler outputs a JSON file that maps each network request to its initiator script and method (step \ding{202}). 
We then label each network request and its initiator code (\eg JS script and methods) as tracking or functional using filter lists \cite{EasyList1, EasyPrivacy} (step \ding{203}).
We use EasyList \cite{EasyList1} and EasyPrivacy \cite{EasyPrivacy} that are used by existing content blockers such as uBlock Origin \cite{ublockOrigin}, Brave \cite{brave}, and Adblock Plus \cite{adblockplusanticv}.
These filter lists only do binary classification and tend to classify mixed resources as functional to avoid website breakage. This is an inherent limitation of filter lists that our work aims to highlight in the context of JavaScript blocking.


\subsubsection{Localizing Tracking and Functional JS Code}
\label{sec:localization}
Next, we classify each script and method using spectra-based ``fault'' localization (SBFL) \cite{sbfl,abreu2009practical}. 
SBFL requires a set of failing and passing test cases. 
For every test, it simply collects the list of code units that participated in the test execution. 
Based on the test output, it labels the participating code units as either passing or failing. 
Finally, it compares the participation of code units in passing and failing tests and assigns a \textit{score} to them.

We adapt SBFL to localize tracking code units (\ie scripts, methods).
Instead of test cases, we analyze each network request and the  methods and scripts in the call stack trace of the network request. 
For example, Figure \ref{fig:annotation}-\ding{202} shows two network requests on {\tt intuit.com}. 
We use filter lists (step \ding{203}) to classify a request (and its call stack) as tracking (\ie failed test case) and functional (\ie passed test case). 
We then calculate ``tracking score'' (Eq \ref{equation: ratio}) for each code unit (\ie script or method) based on its participation in the call stack trace of tracking and functional requests, as shown in step \ding{204}. 
The script {\tt utag.js} initiates 132 tracking requests and 160 functional requests. 
In this script, method {\tt loader} initiates 131 tracking requests and 1 functional request. 
Method {\tt fireCORS} initiated 159 functional and 1 tracking request. 
Figure \ref{fig:annotation} demonstrates the calculation of the tracking score on the webpage in step \ding{205}. 
\begin{table}[t]
\small
    \centering
    \scalebox{1}{\begin{tabular}{ l l l c  c  c}
\hline
\hline
\textbf{ID} & \textbf{Level} & \textbf{JS} &\multicolumn{3}{c}{\bf Blocked Annotated Entity}\\
&&\textbf{block}&\textbf{Tracking}&\textbf{Mixed}&\textbf{Functional}\\
\hline

CTRL & None &None& \ding{55} & \ding{55} &\ding{55}\\
ALL & script &Blanket&  \ding{52} & \ding{52} &\ding{52}\\
TS & script & Selective & \ding{52} & \ding{55} &\ding{55}\\
MS & script & Selective & \ding{55} & \ding{52} &\ding{55}\\
TMS & script  & Track \& Mixed & \ding{52} & \ding{52} &\ding{55}\\
TM & method &Method &  \ding{52} & \ding{55} &\ding{55}\\

\hline
\hline
\end{tabular}}
 \vspace{2ex}
    \caption{Six different JS blocking configurations. \ding{55} represents an unblocked entity, and \ding{52} represents a blocked entity. }
    \label{table:blocking-conf}
\vspace{-8ex}
\end{table}

\begin{equation}
    tracking~score =\log\left(\frac{number~of~tracking~requests}{number~of~functional~requests}\right)
    \label{equation: ratio}
\end{equation}

We classify code units that participate $100\times$ times more in tracking than functional (\ie tracking score of $> 2$ ) as tracking. 
We classify code units that participate $100\times$ times more in functional than tracking (\ie tracking score of $< -2$ ) as functional. 
This threshold is determined experimentally in prior work~\cite{trackersift}. 
The code units that fall in neither category are classified as mixed. 
The localization step results in a list of tracking, functional, and mixed JS methods and scripts. 
In this example, script {\tt utag.js} is classified mixed, method {\tt fireCORS()} is functional, and method {\tt loader()} is tracking.

\subsubsection{JS Blocking Impact Analysis}
To measure the impact of blocking JS code units,  our custom-built Chrome extension loads every page from the 100K websites and blocks the associated tracking JS script or method from the list of labeled methods and scripts. 
It blocks the JS scripts from loading in the browser, similar to existing content blockers.  To block a script method, it simply replaces the method name with {\tt doNotExecuteMe} to redirect its invocations, as shown in Listing \ref{lst:mthd_ex}. Renaming the method name may cause a MethodNotFound exception that terminates the tracking thread in a webpage's JS execution as intended. 

We conduct this experiment on the same 100K webpages in six parallel configurations shown in Table~\ref{table:blocking-conf}. These configurations are illustrated in the {\tt livescore.com} case study and inspired by unique JS blocking strategies that are mostly in practice or proposed by prior work. 
%
Control configuration ({\tt CTRL}) is used to localize JS code units (scripts and methods) using the aforementioned SBFL technique and for breakage comparison in the later subsection.
In {\tt ALL}, all scripts (tracking, mixed, and functional) are blocked to evaluate blanket JS blocking.
 This configuration represents NoScript, which blocks all scripts by default.
%
In {\tt TS}, tracking scripts are blocked to evaluate selective JS blocking.
This configuration represents the majority of content blockers such as uBlock Origin \cite{ublockOrigin}, Brave \cite{brave}, and Adblock Plus \cite{adblockplusanticv} that use EasyList \cite{EasyList1} and EasyPrivacy \cite{EasyPrivacy}.
In {\tt MS}, mixed scripts are blocked to see its adverse consequence on functionality.
In {\tt TMS}, tracking and mixed scripts are blocked to evaluate tracking and mixed JS blocking.
{\tt TMS} is the optimum choice for content blockers in tracking prevention, but it risks functionality breakage, as shown in Section \ref{sec:motivation}.
Finally, we compare the results of {\tt TMS} with {\tt TM}, where we block tracking methods (all located in tracking and mixed scripts) to evaluate method-level JS blocking.
%

%
In {\tt CTRL} configuration, we have websites that do not crash. However, website crashes and breakages may still occur in the blocking configurations due to blocking.
Website breakage is a subjective metric that requires a visual inspection, which is not feasible on 100K webpages. 
Therefore, we discuss two metrics that are correlated with website breakage \cite{le2022autofr}.

\noindent \textbf{Tracking and Functional request count.}
Network requests fetch critical functional resources like scripts, images, and other media as well as JS scripts and images that perform tracking activity. 
We use the number of tracking and functional requests as a measure of tracking and functional activity on a webpage. 
We compare these numbers with the control configuration ({\tt CTRL}) to get the missing requests, as shown in Figure \ref{fig:metrics}-\ding{202}.
This metric helps in collecting non-visual breakage clues.
For example, we do not see any visual breakage on website \textit{poshmark.ca} after blocking mixed script \textit{sdk.js?hash=*}. 
Instead, we observe two missing requests, one that sets the cookie and the other functional request that redirects the login button. 

\noindent \textbf{HTML of websites.}
We scan the HTML tags with {\tt src} attributes on a webpage to estimate visible functional deterioration.
These HTML tags include {\tt <img>}, {\tt <video>}, and {\tt <iframe>}. 
Each tag has a source, {\tt src}, attribute that specifies the URL of a resource file. 
We compare the missing tags in our experiments with the control configuration ({\tt CTRL}), as shown in Figure \ref{fig:metrics}-\ding{203}. 
Note that if the attribute of a missing URL belongs to the functional request in the control configuration ({\tt CTRL}), then it is classified as functional breakage.

\subsection{Phase II:  Manual Inspection of JS Blocking}
\label{sec:study}

\begin{table}
    \small 
    \centering
    \scalebox{1}{\begin{tabular}{ l l c r}
\hline
\hline
\textbf{Script}& \textbf{Script}& \textbf{Method}& \textbf{Websites}\\
\textbf{Domain}&&& \textbf{(\%)}\\
\hline
{\tt google-analytics.com} & {\tt analytics.js} & {\tt wd} & 38\% \\
{\tt google-analytics.com} &{\tt analytics.js}& {\tt ta} & 25\% \\
{\tt facebook.net}&{\tt fbevents.js} & {\tt c}& 19\%\\
{\tt googlesyndication.com}& {\tt sodar2.js} & {\tt Ma}& 11\%\\
{\tt twitter.com}&{\tt widget.js} & {\tt i.e }& 7\%\\
\hline
\hline
\end{tabular}}
    \vspace{3ex}
    \caption{Top JS methods found on the maximum number of websites in control configuration.}
    \label{table:top-methods}
    \vspace{-7ex}
\end{table}
\subsubsection{Data Sampling} 
Manually inspecting 100k websites is time-consuming and practically infeasible. 
We randomly sample 500 websites from the top 100K websites used in Phase I. 
We exclude duplicate websites and websites with the same second-level domains (SLD), but different top-level domains (TLD) \eg {\tt google.com.uk} and {\tt google.com}. 
We excluded a total of 117 websites and manually inspected 383 websites, which is a statistically significant sample size for 100K websites with $\pm$ 5\% margin of error \cite{surveymonkey}. 
\begin{table*}
    \small 
    \centering
    \scalebox{1}{\begin{tabular}{ l r  r  r r r r  r r}
\hline
\hline
 &\multicolumn{3}{c}{\bf Total Network Requests} & \multicolumn{3}{c}{\bf Script-Initiated Network Requests}& \textbf{Total} & \textbf{Total} \\
\textbf{Blocking Configuration}&\textbf{Tracking}&\textbf{Functional}&\textbf{Total}&\textbf{Tracking}&\textbf{Functional}&\textbf{Total}& \textbf{Scripts}&\textbf{JS Methods}\\
\hline
CTRL & 1,175,033 & 4,279,844 & 5,454,877& 953,931 & 882,111 & 1,836,042& 256,042  & 366,025 \\
ALL & 265,101 &  3,248,767 & 3,513,868 & 177,352 & 315,378 & 492,730& 91,984 & 137,006 \\
TS & 355,169  & 4,049,340 & 4,404,509& 248,103 & 820,428 & 1,068,531& 164,670 & 239,960\\
MS & 1,012,708 & 3,916,499 & 4,929,157& 815,553 & 684,084 & 1,499,637& 227,658 & 323,174\\
TMS & 349,888 & 3,887,372 & 4,237,260& 245,389 & 657,361 & 902,750& 155,810 & 224,681 \\
TM & 348,135 & 4,115,351 & 4,463,486& 243,002 & 749,238 & 991,240& 164,543 & 233,927\\
\hline
\hline
\end{tabular}
}
    \vspace{2ex}
    \caption{Characteristics of the crawled dataset across six blocking configurations.}
    \label{table:summary}
    \vspace{-.3 in}
\end{table*}

\begin{figure*}[!t]
\begin{subfigure}{.19\textwidth}
    \begin{tikzpicture}
            \begin{axis}[
                width  = 1.1*\linewidth,,
                height = 4cm,
                name=p2a,
                ybar=2*\pgflinewidth,
                bar width=3pt,
                ylabel = {\% of websites},
                xlabel = {(a) \textbf{Script domains} in {\tt CTRL}},
                symbolic x coords={google-analytics.com, googletagmanager.com, googlesyndication.com, twitter.com, facebook.net, shopify.com,  doubleclick.net, googleapis.com},
                ymin=0,
                ymax=100,
                xtick = data,
                x tick label style={font=\small, rotate = 90},
                xlabel style={text width=3cm, font=\small,align=center},
                ylabel style={text width=3cm, font=\small,align=center},
                legend style={
                    font=\small
            }
            ]
                \addplot[style={last-year,fill=this-year,mark=none}]
                    coordinates {(google-analytics.com, 39.4) (googletagmanager.com,29.5) (googlesyndication.com,28.9)  (twitter.com, 24.7) (facebook.net,24.6 ) (shopify.com, 19.0) (doubleclick.net,15.4) (googleapis.com, 13.1)};
            
            \end{axis}
        \end{tikzpicture}  
    \label{fig:domain-analysis}
    \vspace{-.15in}
\end{subfigure}
\hspace{-0.2in}
\begin{subfigure}{.19\textwidth}
    \begin{tikzpicture}
            \begin{axis}[
                width  = 1.1*\linewidth,,
                height = 4cm,
                name=p2a,
                ybar=2*\pgflinewidth,
                bar width=3pt,
                ymajorgrids = true,
                xlabel = {(b) \textbf{Script domains in} {\tt ALL}},
                symbolic x coords={shopify.com, squarespace.com, googletagmanager.com, parastorage.com, cookiebot.com, ezodn.com, googleapis.com, wp.com},
                ymin=0,
                ymax=100,
                ytick = \empty,
                xtick = data,
                x tick label style={font=\small, rotate = 90},
                xlabel style={text width=3cm, font=\small,align=center},
                ylabel style={text width=1cm, font=\small,align=center},
                legend style={
                    font=\small
            }
            ]
                \addplot[style={last-year,fill=this-year,mark=none}]
                    coordinates { (shopify.com,4.1) (squarespace.com,3.9) (googletagmanager.com,3.4) (parastorage.com,2.6) (cookiebot.com,2.5) (ezodn.com,2.3) (googleapis.com,1.8) (wp.com,1.5)};
            
            \end{axis}
        \end{tikzpicture}
    \label{fig:domain-analysis}
\end{subfigure}
\hspace{-0.4in}
\begin{subfigure}{.19\textwidth}
    \begin{tikzpicture}
            \begin{axis}[
                width  = 1.1*\linewidth,,
                height = 4cm,
                name=p2a,
                ybar=2*\pgflinewidth,
                bar width=3pt,
                ymajorgrids = true,
                xlabel = {(c) \textbf{Script domains in} {\tt TS}},
                symbolic x coords={twitter.com, shopify.com, googleapis.com, tawk.to, google.com, facebook.net, googletagmanager.com, squarespace.com},
                ymin=0,
                ymax=100,
                ytick = \empty,
                xtick = data,
                 xlabel style={text width=3cm, font=\small,align=center},
                x tick label style={font=\small, rotate = 90},
                ylabel style={text width=1cm, font=\small,align=center},
                legend style={
                    font=\small
            }
            ]
                \addplot[style={last-year,fill=this-year,mark=none}]
                    coordinates {(twitter.com,15.0) (shopify.com,12.4) (googleapis.com,11.7) (tawk.to,11.7) (google.com,5.2) (facebook.net,4.6) (googletagmanager.com,4.6) (squarespace.com,3.9)};
            
            \end{axis}
        \end{tikzpicture}
    \label{fig:domain-analysis}
\end{subfigure}
\hspace{-0.4in}
\begin{subfigure}{.19\textwidth}
    \begin{tikzpicture}
            \begin{axis}[
                width  = 1.1*\linewidth,,
                height = 4cm,
                name=p2a,
                ybar=2*\pgflinewidth,
                bar width=3pt,
                ymajorgrids = true,
                xlabel = {(d) \textbf{Script domains in} {\tt MS}},
                symbolic x coords={google-analytics.com, googlesyndication.com, googletagmanager.com, facebook.net, doubleclick.net, googleapis.com, shopify.com, tawk.to},
                ymin=0,
                ymax=100,
                ytick = \empty,
                xtick = data,
                 xlabel style={text width=3cm, font=\small,align=center},
                x tick label style={font=\small, rotate = 90},
                ylabel style={text width=1cm, font=\small,align=center},
                legend style={
                    font=\small
            }
            ]
                \addplot[style={last-year,fill=this-year,mark=none}]
                    coordinates {(google-analytics.com,29.0) (googlesyndication.com,22.5) (googletagmanager.com,20.4) (facebook.net,15.5) (doubleclick.net,12.3) (googleapis.com,9.3) (shopify.com,8.7) (tawk.to,5.5)};
            
            \end{axis}
        \end{tikzpicture}
    \label{fig:domain-analysis}
\end{subfigure}
\hspace{-0.4in}
\begin{subfigure}{.19\textwidth}
    \begin{tikzpicture}
            \begin{axis}[
                width  = 1.1*\linewidth,,
                height = 4cm,
                name=p2a,
                ybar=2*\pgflinewidth,
                bar width=3pt,
                ymajorgrids = true,
                xlabel = {(e) \textbf{Script domains in} {\tt TMS}},
                symbolic x coords={shopify.com, bing.com, squarespace.com, criteo.net, googletagmanager.com, sharethis.com, onesignal.com, ezodn.com},
                ymin=0,
                ymax=100,
                ytick = \empty,
                xtick = data,
                x tick label style={font=\small, rotate = 90},
                xlabel style={text width=3cm, font=\small,align=center},
                ylabel style={text width=1cm, font=\small,align=center},
                legend style={
                    font=\small
            }
            ]
                \addplot[style={last-year,fill=this-year,mark=none}]
                    coordinates {(shopify.com,5.6) (bing.com,5.4) (squarespace.com,4.2) (criteo.net,4.0) (googletagmanager.com,3.8) (sharethis.com,3.6)(onesignal.com,3.6)(ezodn.com,2.8)};
            
            \end{axis}
        \end{tikzpicture}
    \label{fig:domain-analysis}
\end{subfigure}
\hspace{-0.4in}
\begin{subfigure}{.19\textwidth}
    \begin{tikzpicture}
            \begin{axis}[
                width  = 1.1*\linewidth,,
                height = 4cm,
                name=p2a,
                ybar=2*\pgflinewidth,
                bar width=3pt,
                ymajorgrids = true,
                xlabel = {(f) \textbf{Script domains in} {\tt TM}},
                symbolic x coords={shopify.com, googleapis.com, tawk.to, google.com, googletagmanager.com, squarespace.com, facebook.net, gstatic.com},
                ymin=0,
                ymax=100,
                ytick = \empty,
                xtick = data,
                xlabel style={text width=3cm, font=\small,align=center},
                x tick label style={font=\small, rotate = 90},
                ylabel style={text width=1cm, font=\small,align=center},
                legend style={
                    font=\small
            }
            ]
                \addplot[style={last-year,fill=this-year,mark=none}]
                    coordinates {(shopify.com,13.2) (googleapis.com,12.3) (tawk.to,8.6) (google.com,8.1) (googletagmanager.com,6.7)(squarespace.com,5.2)(facebook.net,5.3)(gstatic.com,5.1)};
            
            \end{axis}
        \end{tikzpicture}
    \label{fig:domain-analysis}
\end{subfigure}

\caption{The top domains of request-initiating scripts across six blocking configurations. X-axis shows the top domains of the request-initiating scripts, and Y-axis shows the \% of websites. } 
\vspace{-.1in}
\label{fig:dist-domains}
\end{figure*}
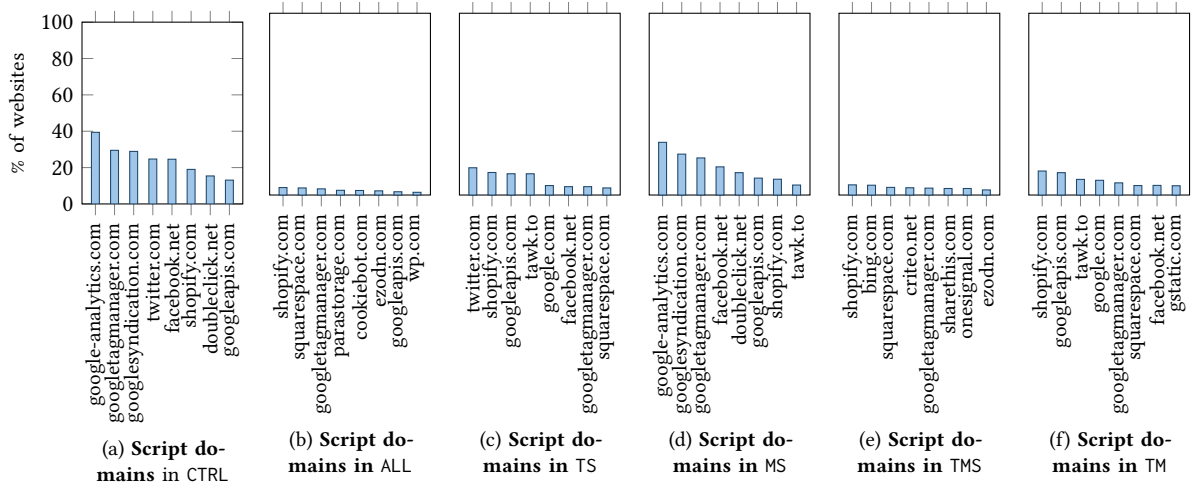 
\subsubsection{Manual Inspection} 
Two testers independently inspected 383 websites.
%
Inspecting six configurations for each website manually and in parallel is prohibitively expensive. 
Therefore, we choose the three most important configurations \ie~{\tt CTRL} (for comparison), {\tt TMS} (tracking and mixed JS blocking), and {\tt TM} (method-level JS blocking).
To assist inspection, our study platform launches three independent instances of Chrome ({\tt CTRL}, {\tt TMS}, and {\tt TM} from Table~\ref{table:blocking-conf}) displayed adjacent to each other. 
Each tester spent at least 5 minutes inspecting the three windows, scrolling each page end to end, and clicking on different webpage components. 
The two testers spent a total of 85 hours manually inspecting the websites and documenting their findings according to the following rubric.
They report visual and functional differences in the following four categories and use a 3-level breakage scale (\ie no breakage, minor breakage, and major breakage).  
Any disagreements were discussed and resolved by consensus.
\begin{itemize}

   \item \textbf{Navigation.} Website navigation contains lists of links to internal webpages. It typically consists of a menu or navigation bar that contains links to various sections of the website, such as the homepage, products or services, about us, and contact. Minor breakage involves non-functional navigation links, abnormal styling layouts, or missing icons. These issues can be frustrating for users and may make it difficult to navigate the website. Major breakage involves more serious issues, such as the navigation button not being operational or the navigation bar not appearing at all. This type of breakage can significantly impact the website's usability. 
   

    \item \textbf{Single sign-on (SSO).} Website SSO allows users to sign in using credentials from services such as Google and Facebook. Minor breakage typically involves issues such as non-functional SSO services, unresponsive login buttons, or missing login options. For example, if the Google SSO service is not functioning, users may be unable to sign in to the website using their Google account. Major breakage involves more serious issues, such as the missing SSO service or the failure of all SSO options. This type of breakage can significantly impact the website's usability.
    

    \item \textbf{Appearance.} This category includes the appearances of media elements, the scrolling behavior of websites, and the HTML element. We exclude advertisements when inspecting appearance-based breakage. Minor breakage involves missing media resources, unstyled HTML, or jittery/unsmooth page scrolling experience. Major breakage involves all the media resources missing altogether or an unscrollable page.
    

    \item \textbf{Additional functionality.} Anything that does not fall into the mentioned categories is added to this category, such as dark mode, website settings, and chatbot.  Minor breakage entails abnormal behavior or non-responsive feature. Major breakage includes page crashes and missing components. 
\end{itemize}

\subsection{Dataset}
This section summarizes the characteristics of dataset crawled across six blocking configurations. 
Table \ref{table:summary} lists the total network requests and script-initiated requests in six configurations over 100K websites and the JS scripts and methods that initiate those requests. %
In control configuration ({\tt CTRL}), out of 5.45 million requests, 22\% of the requests are tracking, leaving the remaining 78\% as functional. 34\% of the total requests are initiated by JS scripts.
In script-initiated requests, 52\% are tracking, and the remaining 48\% are functional.
These script-initiated requests are initiated by 366K JS methods inside 256K scripts. 

Figure \ref{fig:dist-domains} shows the top domains of the scripts that initiate network requests.  
In control configuration ({\tt CTRL}), 39\% of websites initiate requests from the script served by {\tt google-analytics.com}, 
30\% of websites initiate requests from the script served by {\tt googletagma\-nager.com}, and 
29\% of websites initiate requests from the script served by {\tt googlesyndication.com}. 

Our baseline JS blocking configuration is {\tt ALL} in which all tracking, mixed, and functional scripts are blocked. Note that a small number of scripts  may still load in {\tt All}  if such scripts were previously not observed during the localization step in Section \ref{sec:localization}. 
When tracking JS scripts are blocked ({\tt TS} configuration), the majority of tracking script domains disappear, including {\tt google-analyt\-ics.com}.
We observe a relatively lower occurrence of script domains in {\tt TMS} than {\tt TM} because {\tt TMS} blocks all tracking and mixed scripts that include all tracking methods and some functional methods. Whereas in {\tt TM}, only tracking methods are blocked. 
For example, due to the mixed nature of scripts from {\tt facebook.net}, scripts from {\tt facebook.net} appear in {\tt TM}, but not in {\tt TMS}.

\lstdefinestyle{base}{
language=Java,
moredelim=**[is][\color{red}]{@}{@},
moredelim=**[is][\color{codegreen}]{!}{!}
}
\begin{lstlisting}[caption={Methods {\tt wd} and {\tt ta} in {\tt analytics.js} served  by {google-analytics.com} are present on 38\% and 25\% of 100K websites, respectively.},language=Java, label={lst:google-analystics}, style=base]
wd = function(a, b, c, d) {
    var e = O.XMLHttpRequest;
    if (!e) return !1;
    var g = new e;
    if (!("withCredentials" in g)) return !1;
    a = a.replace(/^http:/, "https:");
    g.open("POST", a, !0);
    g.withCredentials = !0;
    g.setRequestHeader("Content-Type", "text/plain");
    g.onreadystatechange = function() {
      if (4 == g.readyState) {
        if (d && "text/plain" === g.getResponseHeader("Content-Type")) try {
          Ea(d, g.responseText, c)
        }
        catch (ca) {
          ge("xhr",
            "rsp"), c()
        } else c();
        g = null}};
    g.send(b);
    return 0}
  ...
ta = function(a) {
    var b = M.createElement("img");
    b.width = 1;
    b.height = 1;
    b.src = a;
    return b}
\end{lstlisting}

Table \ref{table:top-methods} shows the top five request-initiating JS methods across 100k websites. 
Method {\tt wd} in script {\tt analytics.js} is served by {\tt google-analytics.com}. It appears in 38\% of the 100K websites where it sets up a request and its header using XMLHttpRequest \cite{xmlthhtpreq} API, shown in Listing \ref{lst:google-analystics}.  
Method {\tt ta} in script {\tt analytics.js} is served by {\tt google-analytics.com}. It appears in 25\% of the websites where it adds the {\tt <img>} tag with a specific source given as a parameter to the function, shown in Listing \ref{lst:google-analystics}. 
Both of these methods are classified as tracking in the localization step in Section~\ref{sec:localization}.
%
%

\vspace{-2ex}

\section{Results}
\label{sec: results}
This section presents the results of our empirical investigation of different types of JS blocking listed in Table \ref{table:blocking-conf}.

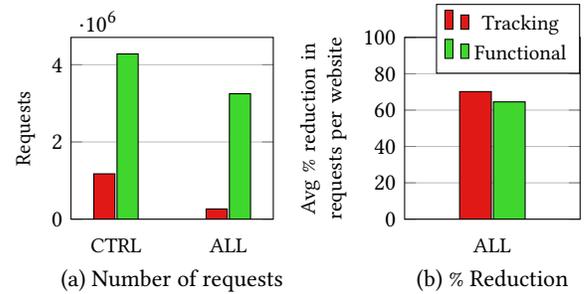
\begin{figure}[t]
        \begin{tikzpicture}
            \begin{axis}[
                width  = 0.24*\textwidth,
                height = 4cm,
                name=p2a,
                major x tick style = transparent,
                ybar=2*\pgflinewidth,
                ymin=0,
                bar width=8pt,
                enlarge x limits=0.4,
                ymajorgrids = true,
                ylabel = {Requests},
                xlabel={(a) Number of requests},
                symbolic x coords={CTRL, ALL},
                xtick = data,
                scaled y ticks = true,
                x tick label style={font=\small,text width=1.5cm,align=center},
                ylabel style={text width=1cm, font=\small ,align=center},
                legend style={
                    at={(1,1)},
                    font=\small,
                    anchor= south east,
            },
            legend columns = 2,
            ]
                \addplot[style={fill=rred,mark=none}]
                    coordinates {(CTRL,1175033) (ALL,265101)};
            
                \addplot[style={fill=ggreen,mark=none}]
                    coordinates {(CTRL,4279844) (ALL,3248767)};
            \end{axis}
            \begin{axis}[
                width  = 0.22*\textwidth,
                height = 4cm,
                name=p3a,
                at=(p2a.right of south east),
                anchor=left of south west,
                major x tick style = transparent,
                ybar=2*\pgflinewidth,
                bar width=12pt,
                enlarge x limits=0.2,
                ymajorgrids = true,
                ylabel = {Avg \% reduction in requests per website},
                xlabel={(b) \% Reduction},
                symbolic x coords={ALL},
                xtick = data,
                ymin=0,
                ymax=100,
                scaled y ticks = true,
                x tick label style={font=\small,text width=2cm,align=center},
                ylabel style={text width=3cm, align=center, font=\small},
                legend style={
                    at={(1,0.8)},
                    font=\small,
                    anchor= south east,
            }
            ]
                \addplot[style={fill=rred,mark=none}]
                    coordinates {(ALL,70.18)};
            
                \addplot[style={fill=ggreen,mark=none}]
                    coordinates {(ALL,64.53)};
                \legend{Tracking,Functional}
            \end{axis}
        \end{tikzpicture}
        \caption{(a) compares the request count of control configuration with blanket JS blocking ({\tt ALL}). (b) shows average \% reduction in request per website for blanket JS blocking ({\tt ALL}).}
      \label{fig:rq1_req}
\end{figure} 
\begin{figure}[t]
       \begin{tikzpicture}
          \begin{axis}[
                ybar=1*\pgflinewidth,
                bar width=6pt,
                height = 4.5cm,
                width=0.49*\textwidth,
                ymajorgrids = true,
                ylabel = {Websites},
                ymin=0,
                ymax=50000,
                xlabel = {\% of blocked requests ({\tt ALL})},
                symbolic x coords={0-10, 11-20, 21-30, 31-40, 41-50, 51-60, 61-70, 71-80, 81-90, 91-100},
                xtick = data,
                scaled y ticks = true,
                x tick label style={font=\small,rotate=90},
                ylabel style={text width=1cm, font=\small ,align=center},
                legend style={
                    at={(0.98,0.98)},
                    font=\small
            }
            ]
                \addplot[style={fill=rred,mark=none}]
                    coordinates {(0-10,34289) (11-20,3229) (21-30,3466) (31-40,4446) (41-50,7969)(51-60,4367)(61-70,5940)(71-80,6696)(81-90,6305) (91-100,23293)};
            
                \addplot[style={fill=ggreen,mark=none}]
                    coordinates {(0-10,31394) (11-20,12066) (21-30,7618) (31-40,6812) (41-50,4833)(51-60,4110)(61-70,3859)(71-80,3879)(81-90,5026) (91-100,20403)};
            
                \legend{Tracking,Functional}
            \end{axis}   
       \end{tikzpicture}
        \caption{The \% of blocked request in blanket JS blocking configuration ({\tt ALL}). Low \% of blocked functional requests and high \% of blocked tracking requests are desirable.}
      \label{fig:rq1_cdf}
\end{figure}
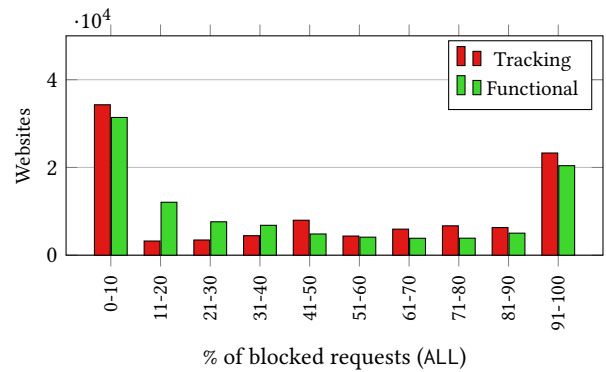 
\begin{table}
    \centering
    \scalebox{1}{\begin{tabular}{ l  r }
\hline
\hline
\textbf{Tag}& \textbf{Blanket JS}\\
\textbf{Category}& \textbf{Blocking (ALL)}\\
\hline
{\tt <image>} & 70600 \\
{\tt <video>} & 5\\
{\tt <iframe>} & 21052\\
{\tt <script>} & 100278\\
{\tt <source>} & 39\\
\hline
\hline
\end{tabular}}
    \vspace{2ex}
    \caption{Missing HTML tags whose URLs are classified as functional in blanket JS blocking ({\tt ALL}).}
    \label{table:rq1}
    \vspace{-4ex}
\end{table}
\subsection{Phase I: Large-scale JS Blocking Analysis}
We aim to address the following research questions in our analysis of JS blocking. 

\begin{enumerate}
\item How resilient is website functionality against blanket JS blocking ({\tt ALL})?

\item How effective is selective script-level JS blocking in tracking prevention and functionality preservation ({\tt TS} and {\tt MS})?

\item How common is it for website developers to mix tracking and functionality in the same script?

\item How effective is method-level JS blocking in tracking prevention and functionality preservation ({\tt TMS} and {\tt TM})?

\end{enumerate}


\subsubsection{RQ1: Blanket JS Blocking}
We first study the naive approach to JS blocking by blocking all JS scripts ({\tt ALL} configuration in Table \ref{table:blocking-conf}). 
Specifically, we block all 256K scripts on 100K webpages and compare the breakage metrics (\ie network request count and HTML resource count) with the control ({\tt CTRL}). 
Given blanket JS blocking, we expect a sharp drop in the number of tracking or functional requests. 
Figure \ref{fig:rq1_req} (a) shows that 22\% of functional requests and 76\% of tracking requests remain after blocking all JS scripts ({\tt ALL}). 
Note that a few requests are initiated by the scripts previously not captured in the localization step in Section \ref{sec:localization} and hence, were not blocked in blanket JS blocking ({\tt ALL}) configuration.
Figure~\ref{fig:rq1_req} (b) presents the average percentage of reduction in request count per webpage. 
On average, per webpage, the tracking and functional request count decrease by 70\% and 65\%, respectively.
This shows that webpages today can retain one-third of functionality even with extreme blocking strategies.
Another observation is that the tracking reduction per webpage is higher than functional reduction, which means that many webpages often sacrifice tracking but attempt to retain functionality.  


\begin{figure}[t]
        \begin{tikzpicture}
            \begin{axis}[
                width  = 0.24*\textwidth,
                height = 4cm,
                name=p2a,
                major x tick style = transparent,
                ybar=2*\pgflinewidth,,
                ymin=0,
                bar width=6pt,
                 enlarge x limits=0.2,
                ymajorgrids = true,
                ylabel = {Requests},
                xlabel={(a) Number of requests},
                symbolic x coords={CTRL, TS, MS},
                xtick = data,
                scaled y ticks = true,
                x tick label style={font=\small,text width=1cm,align=center},
                ylabel style={text width=1cm, font=\small ,align=center},
                legend style={
                    at={(1,1.0)},
                    font=\small,
                    anchor= south east,
            }
            ]
                \addplot[style={fill=rred,mark=none}]
                    coordinates {(CTRL,1175033) (TS,355169) (MS,1012708)};
            
                \addplot[style={fill=ggreen,mark=none}]
                    coordinates {(CTRL,4279844) (TS,4049340) (MS,3916499)};
            
            \end{axis}
            \begin{axis}[
                width  = 0.22*\textwidth,
                height = 4cm,
                name=p3a,
                at=(p2a.right of south east),
                anchor=left of south west,
                major x tick style = transparent,
                enlarge x limits=0.5,
                ybar=1*\pgflinewidth,,
                bar width=8pt,
                ymajorgrids = true,
                ylabel = {Avg \% reduction in requests per website},
                symbolic x coords={TS, MS},
                xlabel={(b) \% Reduction},
                xtick = data,
                ymin=0,
                ymax=100,
                scaled y ticks = false,
                x tick label style={font=\small,text width=1cm,align=center},
                ylabel style={text width=3cm,align=center,  font=\small},
                legend style={
                    at={(1,0.8)},
                    font=\small,
                    anchor= south east,
            }
            ]
                \addplot[style={fill=rred,mark=none}]
                    coordinates { (TS,57.02) (MS,12.24)};
            
                \addplot[style={fill=ggreen,mark=none}]
                    coordinates { (TS,11.76) (MS,19.83)};
                    
                \legend{Tracking,Functional}
            \end{axis}
        \end{tikzpicture}
        \caption{(a) compares the request count of control configuration with selective JS blocking ({\tt TS} and {\tt MS}). (b) shows average \% reduction in request per website for selective JS blocking ({\tt TS} and {\tt MS}).}
      \label{fig:rq3_req}

\end{figure}
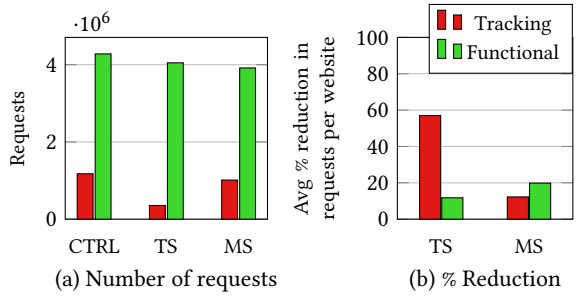
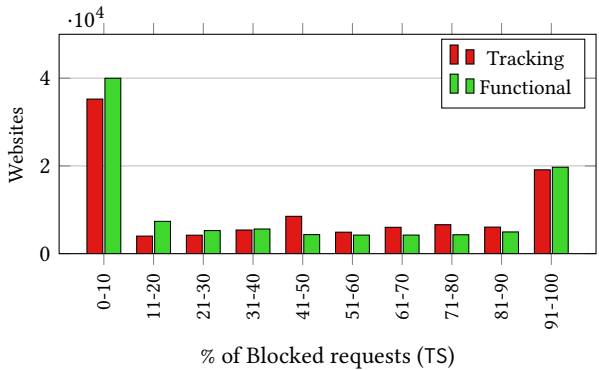
\begin{figure}[t]
       \begin{tikzpicture}
           \begin{axis}[
                ybar=2*\pgflinewidth,
                bar width=6pt,
                height = 4.5cm,
                width=0.49*\textwidth,
                ylabel = {Websites},
                ymin=0,
                ymax=50000,
                ymajorgrids = true,
                xlabel = {\% of Blocked requests ({\tt TS})},
                symbolic x coords={0-10, 11-20, 21-30, 31-40, 41-50, 51-60, 61-70, 71-80, 81-90, 91-100},
                xtick = data,
                scaled y ticks = true,
                x tick label style={font=\small,rotate=90},
                ylabel style={text width=1cm, font=\small ,align=center},
                legend style={
                    at={(0.98,0.98)},
                    font=\small,
            }
            ]
                \addplot[style={fill=rred,mark=none}]
                    coordinates {(0-10,35235) (11-20,4001) (21-30,4206) (31-40,5389) (41-50,8509)(51-60,4892)(61-70,6006)(71-80,6603)(81-90,6051) (91-100,19108)};
            
                \addplot[style={fill=ggreen,mark=none}]
                    coordinates {(0-10,39983) (11-20,7362) (21-30,5262) (31-40,5618) (41-50,4346)(51-60,4232)(61-70,4233)(71-80,4311)(81-90,4950) (91-100,19703)};
            
                \legend{Tracking,Functional}
            \end{axis}
        \end{tikzpicture}
    \vspace{-1ex}
        \caption{The \% of blocked request in selective JS blocking configuration ({\tt TS}). Low \% of blocked functional requests and higher \% of blocked tracking requests are desirable.}
      \label{fig:rq2_cdf}
     \vspace{-1ex}
\end{figure} 
\begin{table}

    \centering
    \scalebox{1}{\begin{tabular}{ l  r  r }
\hline
\hline
\textbf{Tag}& \textbf{Tracking}& \textbf{Mixed}\\
\textbf{Category}& \textbf{JS Blocked (TS)}& \textbf{JS Blocked (MS)}\\

\hline

{\tt <image>} & 12607 & 20035\\
{\tt <video>} & 0& 0\\
{\tt <iframe>} & 11774& 14682\\
{\tt <script>} & 21650& 37197\\
{\tt <source>} & 23& 37\\

\hline
\hline
\end{tabular}}
    \vspace{2ex}
    \caption{Missing HTML tags whose URLs are classified as functional in selective JS blocking ({\tt TS} and {\tt MS}) .}
    \label{table:rq2}
    \vspace{-7ex}
\end{table}

To map this behavior per webpage, we find the number of webpages with different levels of request reduction for both tracking and function. 
Figure \ref{fig:rq1_cdf} illustrates the result.
We find that the majority of the webpages (57\%) have either less than 10\% request reduction or more than 90\% request reduction in both tracking and functional.
This result shows both (1) high resilience against tracking reduction and functional breakage due to anti-content blocking strategies such as loading resources by changing network endpoints \cite{Le21anticvndss,Alrizah19IMCerrorsMisunderstandings}, and also (2) low resilience where blocked scripts are critical for a functioning webpage \cite{sugarcoat,trackersift}. 
Further inspection of HTML DOM elements reveal that 191K functional HTML tag sources are missing from 100K webpages when {\tt ALL} scripts are blocked, reflecting severe functionality loss. 
Table \ref{table:rq1} shows the breakdown of the category of these missing sources.
In {\tt ALL} configuration, 71K functional {\tt <img>} tags, 21K functional {\tt <iframe>} tags, and 100K functional {\tt <script>} tags are missing.

\begin{tcolorbox}
{
{\bf \em Takeaway.} Two-thirds (66\%) of the webpages experience a significant functionality breakage when blanket JS blocking is employed. 
}
\end{tcolorbox}

\subsubsection{RQ2: Effectiveness of Selective JS Blocking}
Since Blanket JS blocking is ineffective, we study the effectiveness of selective JS blocking by blocking tracking scripts ({\tt TS} configuration in Table \ref{table:blocking-conf}). Later, we block mixed scripts ({\tt MS} configuration in Table \ref{table:blocking-conf}) to see its adverse effects on functionality.

\noindent\textbf{Blocking Tracking Scripts.} In this experiment, we block 93K tracking scripts ({\tt TS}) from 256K JS scripts across 100K live webpages and investigate its impact on tracking mitigation and functional breakage. 
Figure \ref{fig:rq3_req} (a) reports that 95\% of functional requests persist, whereas  30\% of tracking requests manage to survive. 
Figure~\ref{fig:rq3_req} (b) shows an average reduction in requests per webpage. In the case of {\tt TS}, we observe a 57\% reduction in tracking requests and an 11\% reduction in functional requests per webpage on average. 
Measurement with HTML tag metric in Table \ref{table:rq2} shows that blocking tracking JS scripts ({\tt TS}) results in 46K missing functional sources across 100K webpages. 
In {\tt TS} configuration, 13K functional {\tt <img>} tags, 12K functional {\tt <iframe>} tags, and 22K functional {\tt <script>} tags are missing.

\begin{figure}[t]
    \centering
        \includegraphics[width=0.49 \textwidth]{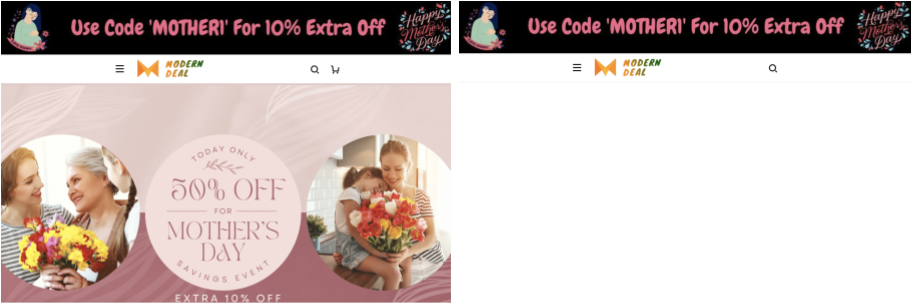} 
         \begin{tikzpicture}[overlay]
          \draw[red,thick,rounded corners] (-4.1,0.5) rectangle (0,2.5);
          \draw[red,thick,rounded corners] (0.3,0.5) rectangle (4.3,2.5);
        \end{tikzpicture}
        \vspace{-3ex}
        \caption{Visual impact of blocking mixed JS script. The left side shows a normal website, whereas the right side shows a breakage due to blocking.}
        \label{fig:pressl}
\end{figure}
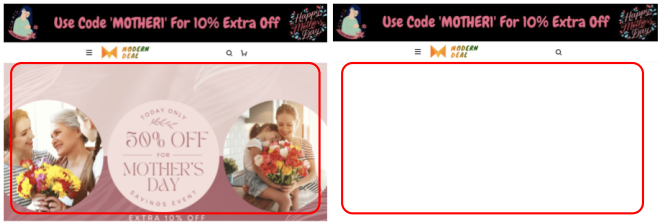

\noindent\textbf{Blocking Mixed Scripts.}
In this experiment, we block only mixed JS scripts ({\tt MS}). 
We expect a decrease in both functionality and tracking, as mixed scripts represent both. 
Figure \ref{fig:rq3_req} (a) visualizes these results. Overall, we see 86\% of tracking and 92\% of functional requests. 
This observation is consistent with other HTML tag metric in Table \ref{table:rq2}. 
In {\tt MS} configuration, 20K functional {\tt <img>} tags, 15K functional {\tt <iframe>} tags, and 37K functional {\tt <script>} tags are missing.
Figure \ref{fig:pressl} show visual breakage on {\tt pressl.co} due to blocking mixed JS scripts that eliminate tracking at the cost of critical functional breakage. 

We further ask {\em Do all webpages react similarly when tracking scripts are blocked?} 
Our goal is to unfold the resilience of different webpages with blocked tracking scripts ({\tt TS}). 
Figure \ref{fig:rq2_cdf} measures the distribution of webpages across different levels of functional breakage and tracking mitigation from blocking tracking scripts. 
39K webpages experience less than 10\% functional deterioration, and  35K webpages experience less than 10\% tracking mitigation. 
The left of the bar chart represents webpages that heavily employ mixed scripts, making JS script blocking ineffective. 
19K webpages are only left with greater than 90\% functionality deterioration and tracking mitigation, representing the class of webpages relying less on mixing scripts and thus are susceptible to JS script blocking.
Although JS script blocking is effective on a few webpages, it does not apply to a significant proportion of webpages that employ mixed scripts. 
Therefore, we must address the tracking behavior concealed in mixed scripts.

\begin{tcolorbox}
{{\bf \em Takeaway.} To maximize tracking prevention while minimizing functional breakage, mixed scripts need to be inspected at a finer granularity.  }
\end{tcolorbox}

\subsubsection{RQ3: Prevalence of Mixed Scripts}
A trivial way for web developers and trackers to bypass filter lists is by mixing functional behavior with tracking in a single script. 
Privacy-enhancing content blockers, such as uBlock Origin, cannot afford to break the webpage and have no choice but to allow such scripts to load in the browser.
To gather concrete evidence on the prevalence of this practice, we first conduct a longitudinal experiment on the frequency of mixed JS scripts over the past two years (2021 and 2022) on 100K webpages. 
%
%
In 2021, we crawled 100K webpages and classified the collected JS code using the SBFL-inspired approach from Section \ref{sec:methodology}. 
We repeat the same experiment in 2022 on the same 100K webpages.

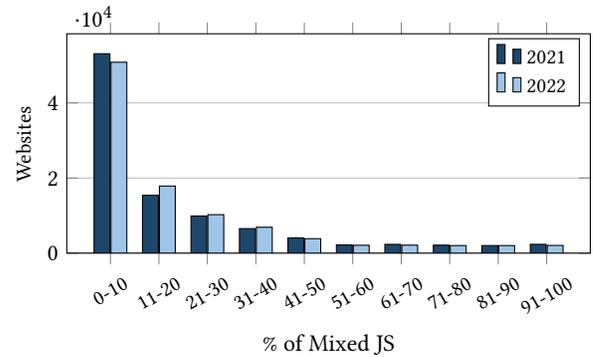
\begin{figure}[t]
        \begin{tikzpicture}
            \begin{axis}[
                width  = 0.48*\textwidth,
                height = 4.5cm,
                name=p2a,
                ybar=1*\pgflinewidth,
                bar width=6pt,
                ymajorgrids = true,
                ylabel = {Websites},
                xlabel = {\% of Mixed JS},
                symbolic x coords={0-10, 11-20, 21-30, 31-40, 41-50, 51-60, 61-70, 71-80, 81-90, 91-100},
                ymin=0,
                xtick = data,
                scaled y ticks = true,
                x tick label style={font=\small, rotate=30},
                ylabel style={text width=1cm, font=\small ,align=center},
                legend style={
                    font=\small
            }
            ]
                \addplot[style={fill=last-year,mark=none}]
                    coordinates {(0-10,53095) (11-20,15413) (21-30,9891) (31-40,6528) (41-50,4046)(51-60,2193)(61-70,2338)(71-80,2129)(81-90,2019) (91-100,2348)};
            
                \addplot[style={fill=this-year,mark=none}]
                    coordinates {(0-10,50828) (11-20,17863) (21-30,10243) (31-40,6899) (41-50,3851)(51-60,2098)(61-70,2128)(71-80,2018)(81-90,2017) (91-100,2055)};
            
                \legend{2021,2022}
            \end{axis}
        \end{tikzpicture}
    \vspace{-1ex}
        \caption{Comparison of \% mixed JS scripts when tracking score is in [-2,2] for web corpus collected in 2021 and 2022.}
      \label{fig:rq2_year}
     \vspace{-2ex}
\end{figure} 
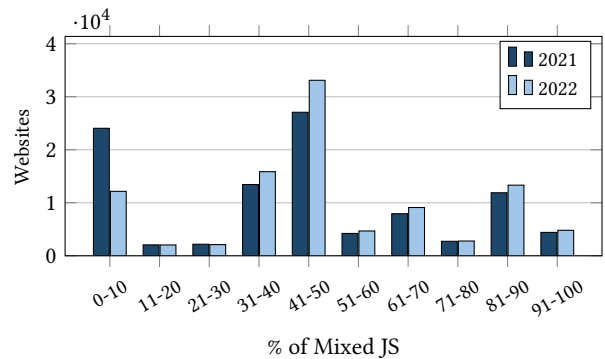
\begin{figure}[t]
        \begin{tikzpicture}
            \begin{axis}[
                width  = 0.49*\textwidth,
                height = 4.5cm,
                name=p3a,
                name=p2a,
                at=(p2a.right of south east),
                anchor=left of south west,
                ybar=1*\pgflinewidth,
                bar width=6pt,
                enlarge y limits={0.25,upper},
                ymajorgrids = true,
                ylabel = {Websites},
                xlabel = {\% of Mixed JS},
                symbolic x coords={0-10, 11-20, 21-30, 31-40, 41-50, 51-60, 61-70, 71-80, 81-90, 91-100},
                xtick = data,
                scaled y ticks = true,
                x tick label style={font=\small,rotate=30},
                ylabel style={text width=1cm, font=\small ,align=center},
                ymin = 0,
                legend style={
                    font=\small
            }
            ]
                \addplot[style={fill=last-year,mark=none}]
                    coordinates {(0-10,24054) (11-20,2059) (21-30,2179) (31-40,13444) (41-50,27079)(51-60,4215)(61-70,7936)(71-80,2732)(81-90,11888) (91-100,4414)};
            
                \addplot[style={fill=this-year,mark=none}]
                    coordinates {(0-10,12159) (11-20,2046) (21-30,2112) (31-40,15870) (41-50,33110)(51-60,4678)(61-70,9109)(71-80,2785)(81-90,13325) (91-100,4806)};
            
                \legend{2021,2022}
            \end{axis}
        \end{tikzpicture}
    \vspace{-1ex}
        \caption{Comparison of \% mixed JS scripts without any threshold on tracking scores for web corpus in 2021 and 2022. }
      \label{fig:rq3_year}
\end{figure}

Figure \ref{fig:rq2_year} shows the result of the experiment. 
The x-axis represents the percentage of scripts that are mixed, ranging from 0 to 100 in 10 bins each of size 10. %
The y-axis represents the number of webpages in each bin. 
In 2021, 15\% of webpages out of 100K have between 11\% to 20\% of scripts that were mixed. 
This number increases to 18\% in 2022. 
Overall, in 2021, out of 220K JS scripts, 28K are mixed JS scripts, making it 12.8\%, whereas, in 2022, 37.5K out of 256K JS scripts are mixed, making it 14.6\%. 
There is 14\% increase in the number of websites employing mixed scripts over 100K websites, as compared to last year.
%
For example, on the website {\tt kixie.com}, we observe a new mixed JS script {\tt 20564323.js} in 2022, initiating HubSpot analytics code along with the functional code that redirects the {\tt Try Kixie Free} button. 
We also find that the change in total script count corroborates the general belief that JS scripts across the web have increased marginally since 2021 \cite{goel_2022}.

While investigating selective JS blocking, we also find deterioration in the functionality when only tracking scripts are blocked ({\tt TS}). 
Naturally, we ask {\em why does blocking tracking scripts (TS) result in functional deterioration?} 
We suspect that such an issue may arise due to the narrow threshold on SBFL's tracking score. JS code units (\ie scripts, methods) with $> 2$ score are annotated as purely tracking. Functional behavior in tracking scripts can also exist due to the dynamic nature of webpages. Between the tracking score measurement and blocking experiments, the script may have changed, or the webpage deliberately refactors the script slightly for reasons such as JS obfuscation \cite{Skolka19minifiedobfuscatedJS,ngan2022nowhere} or minification \cite{moog2021statically}. For better threshold selection,  we must answer {\em what are the consequences of widening the tracking score threshold?} We conduct a brief sensitivity analysis on the tracking score's threshold. 
Figure \ref{fig:rq3_year} shows the new distribution when the threshold is set to maximum. 
We find that 46\% of the webpages have more than 50\% of their scripts mixed with at least one tracking or functional request, further reducing the applicability of JS script blocking and showing the extent of this problem. Our investigation in RQ3 highlights the following trade-off. We either sacrifice functionality when blocking mixed JS scripts or let go of privacy. If functional preservation is critical, we forego opportunities to block numerous tracking activities.



\begin{tcolorbox}
{{\bf \em Takeaway.} Websites are increasingly employing sophisticated code refactoring techniques (\eg inlining or bundling) to mix tracking code with functional code, making existing content-blocking techniques ineffective.}
\end{tcolorbox}
  

\subsubsection{RQ4: Fine-Grained JS Blocking}
In RQ4, we assess the benefits of performing JS blocking at the method-level. 
Our hypothesis is that blocking tracking JS method will provide higher precision in tracking prevention, leading to significantly lower functional breakage than JS script-level blocking. 
In our first experiment, we compare the effectiveness of method-level JS blocking ({\tt TM}) against tracking and mixed JS blocking ({\tt TMS}).

We combine results from blocking both tracking and mixed scripts ({\tt TMS}) as the baseline because all tracking methods are either located in tracking scripts or mixed scripts. 
Blocking a tracking JS method ({\tt TM}) may eliminate the tracking behavior of a mixed script or a tracking script.

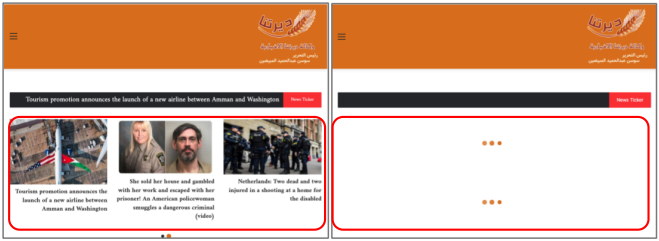
\begin{figure}[t]
    \centering
        \includegraphics[width=0.49 \textwidth]{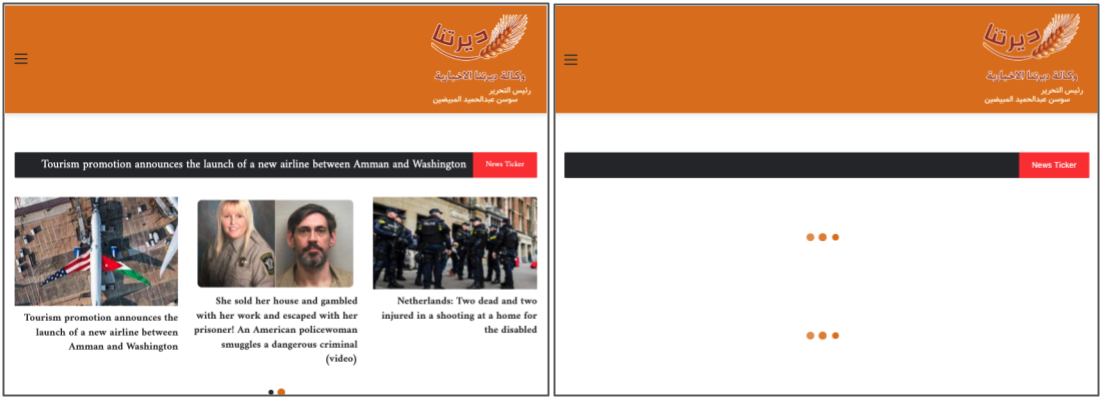} 
         \begin{tikzpicture}[overlay]
          \draw[red,thick,rounded corners] (-4.1,0.5) rectangle (0.1,2.0);
          \draw[red,thick,rounded corners] (0.2,0.5) rectangle (4.4,2.0);
        \end{tikzpicture}
        \vspace{-3ex}
        \caption{Image compares the functional breakage in  tracking and mixed JS blocking (right) as compared to method-level JS blocking (left), which loads the website {\tt deeretnanews.com} normally.}
        \label{fig:news}
\end{figure}

Figure \ref{fig:rq4_req} summarizes these results. Both baseline tracking and mixed JS blocking ({\tt TMS}) and method-level JS blocking ({\tt TM}) reduce the tracking requests by 71\% and block on average 62\% of the tracking requests per page. The two configurations cover most of the tracking requests among themselves, and blocking them will yield the same result. 
More surprisingly, we see an improvement in total functionality retention when blocking method-level ({\tt TM}) \ie a 6\% total improvement, whereas the average functional request breakage per page decreases by 7\%.  
 On evaluating HTML, JS method-level blocking({\tt TM}) retains approximately 2X more functional HTML tag sources, such as images and scripts, than blocking tracking and mixed JS scripts ({\tt TMS}), as shown in Table \ref{table:rq4}. 
 For example, in Figure \ref{fig:news}, we visually inspect {\tt deeretnanews.com} to find functional media breakage in {\tt TMS} configuration that loads normally in {\tt TM} configuration.

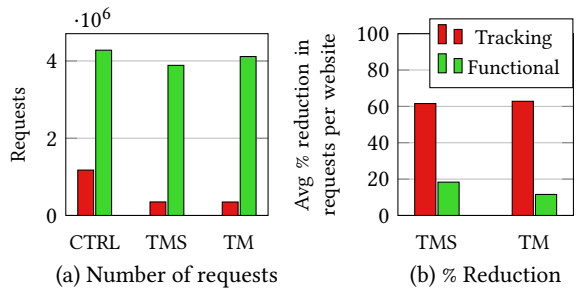
\begin{figure}[t]
        \begin{tikzpicture}
            \begin{axis}[
                width  = 0.24*\textwidth,
                height = 4cm,
                name=p2a,
                major x tick style = transparent,
                ybar=2*\pgflinewidth,,
                ymin=0,
                bar width=6pt,
                 enlarge x limits=0.2,
                ymajorgrids = true,
                ylabel = {Requests},
                symbolic x coords={CTRL, TMS, TM},
                xtick = data,
                scaled y ticks = true,
                xlabel={(a) Number of requests},
                x tick label style={font=\small,text width=1cm,align=center},
                ylabel style={text width=1cm, font=\small ,align=center},
                legend style={
                    at={(1.5,1)},
                    font=\small,
                    anchor= south east,
            },     legend columns=2,
            ]
                \addplot[style={fill=rred,mark=none}]
                    coordinates {(CTRL,1175033) (TMS,349888) (TM,348135)};
            
                \addplot[style={fill=ggreen,mark=none}]
                    coordinates {(CTRL,4279844) (TMS,3887372) (TM,4115351)};
            
            \end{axis}
            \begin{axis}[
                width  = 0.22*\textwidth,
                height = 4cm,
                name=p3a,
                at=(p2a.right of south east),
                anchor=left of south west,
                major x tick style = transparent,
                enlarge x limits=0.2,
                ybar=2*\pgflinewidth,,
                bar width=8pt,
                enlarge x limits=0.4,
                ymajorgrids = true,
                ylabel = {Avg \% reduction in requests per website},
                symbolic x coords={TMS, TM},
                xlabel={(b) \% Reduction},
                xtick = data,
                ymin=0,
                ymax=100,
                scaled y ticks = false,
                x tick label style={font=\small,text width=1cm,align=center},
                ylabel style={text width=3cm,align=center,  font=\small},
                legend style={
                    at={(1,0.7)},
                    font=\small,
                    anchor= south east,
            }
            ]
                \addplot[style={fill=rred,mark=none}]
                    coordinates {(TMS,61.56) (TM,62.81)};
            
                \addplot[style={fill=ggreen,mark=none}]
                    coordinates {(TMS,18.32) (TM,11.56)};
                    
                \legend{Tracking,Functional}
            \end{axis}
        \end{tikzpicture}

        \caption{(a) compares the request count of control configuration with tracking and mixed ({\tt TMS}) and method-level JS blocking ({\tt TM}). (b) shows average \% reduction in request per website for tracking and mixed ({\tt TMS}) and method-level JS blocking ({\tt TM}).}
      \label{fig:rq4_req}
\vspace{-.1in}
\end{figure} 
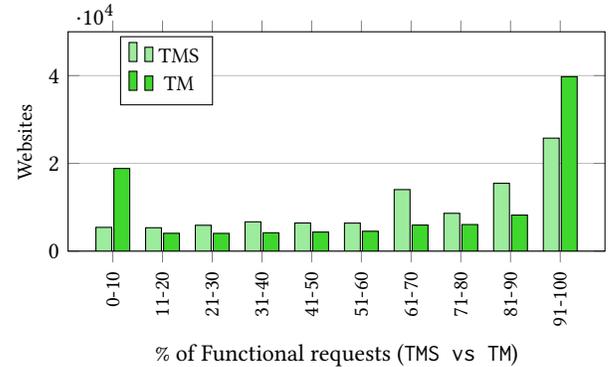
\begin{figure}[t]
       \begin{tikzpicture}
            \begin{axis}[
                width  = 0.49*\textwidth,
                height = 4.5cm,
                ybar=2*\pgflinewidth,
                bar width=6pt,
                ymajorgrids = true,
                ymin=0,
                ymax=50000,
                xlabel = {\% of Functional requests ({\tt TMS vs TM})},
                symbolic x coords={0-10, 11-20, 21-30, 31-40, 41-50, 51-60, 61-70, 71-80, 81-90, 91-100},
                xtick = data,
                ylabel = {Websites},
                scaled y ticks = true,
                x tick label style={font=\small,rotate=90},
                ylabel style={text width=1cm, font=\small ,align=center},
                legend style={
                    at={(0.27, 0.98)},
                    font=\small
            }
            ]
                \addplot[style={fill=ggreen1,mark=dotted}]
                    coordinates {(0-10,5419) (11-20,5332) (21-30,5895) (31-40,6681) (41-50,6416)(51-60,6390)(61-70,14027)(71-80,8621)(81-90,15464) (91-100,25755)};
            
                \addplot[style={fill=ggreen,mark=none}]
                    coordinates {(0-10,18841) (11-20,4068) (21-30,4047) (31-40,4187) (41-50,4344)(51-60,4532)(61-70,5942)(71-80,6060)(81-90,8217) (91-100,39762)};
            
                \legend{TMS,TM}
            \end{axis}
        \end{tikzpicture}
        \caption{The \% of functional requests in tracking and mixed ({\tt TMS}) JS blocking and method-level JS blocking ({\tt TM}). A higher \% of functional requests is desirable.}
      \label{fig:rq4_cdf}
\end{figure} 
\begin{table}
    \centering
    \scalebox{1}{\begin{tabular}{l  r  r }
\hline
\hline
\textbf{Tag}& \textbf{Tracking \& Mixed}& \textbf{Tracking JS Methods}\\
\textbf{Category}& \textbf{JS Blocked (TMS)}& \textbf{Blocked (TM)}\\

\hline
\hline
{\tt <image>} & 30512 & 17524 \\
{\tt <video>} & 0& 2 \\
{\tt <iframe>} & 18362 & 14035 \\
{\tt <script>} & 56852& 30011 \\
{\tt <source>} & 37& 35 \\

\hline
\hline
\end{tabular}}
    \vspace{2ex}
    \caption{Missing HTML tags whose URLs are classified as functional in tracking and mixed ({\tt TMS}) and method-level ({\tt TM}) JS blocking.}
    \label{table:rq4}
    \vspace{-.3in}
\end{table}

We further investigate {\em  how much functional breakage does each webpage face with method-level blocking ({\tt TM}) compared to the baseline {\tt TMS}?}
Figure \ref{fig:rq4_cdf} sheds more light on the functional request count between two blocking granularities. With method-level JS blocking ({\tt TM}), 40\% webpages have less than 10\% functional breakage (preserved more than 90\% functional requests). In comparison, tracking and mixed JS blocking ({\tt TMS}) leads to around 25\% of webpages in this category. 

We observe two classes of webpages: (1) webpages that decouple functionality and tracking more prominently at the method-level and hence, are less prone to functional breakage, and (2) webpages that tightly integrate tracking code with functional, which is harder to separate even at the method-level and thus results in high functional breakage when such methods are blocked. Further investigation on the number of such mixed methods finds that  6\% of 366k JS methods integrate tracking with functional code.

\begin{tcolorbox}
{\bf \em Takeaway.} Nearly 40\% of the webpages implement functional and tracking code in a modularized fashion. 
Blocking tracking methods in such webpages shows improved tracking prevention and reduced functional breakage as compared to script-level blocking. 
The rest of the webpages demand increasing the granularity  (\ie statement-level) or incorporating more sophisticated dynamic analysis. 
\end{tcolorbox}

\subsection{Phase II: Visual Inspection of JS Blocking and Web Breakage} 
In Phase II, we perform a qualitative study to validate our quantitative findings with an in-depth visual inspection of sampled websites, as described in Section~\ref{sec:study}. We seek to answer the following research questions: 

\begin{enumerate}[resume]

\item Does our manual inspection validate that method-level JS blocking is more effective than JS blocking?


\item Is method-level JS blocking the most effective in minimizing breakage while preventing tracking?

\item Can webpages withstand the removal of tracking methods?

\end{enumerate}

\subsubsection{RQ5: Validating the effectiveness of method-level JS blocking.}

Figure \ref{fig:minor} and Figure \ref{fig:major} summarizes the results of investigating true functional breakage on 383 websites, measured according to four established metrics (\ie navigation, SSO, appearance, and others) and three levels of breakage. The X-axis represents the percentage of websites with functional breakage. Overall, there is an evident decline in the number of broken websites, for both major and minor breakage, when JS method-level blocking is used instead of tracking and mixed JS blocking. These results validate the findings of quantitative analysis in RQ4. In tracking and mixed JS blocking ({\tt TMS}), 68 websites have minor breakage and 118 websites have major breakage, whereas, in method-level JS blocking, 45 websites have minor breakage, and 29 websites have major breakage. Most of the breakages were observed in additional feature categories, comprising broken widgets (\eg chatbots and feedback) and malfunctioning home buttons.

{\tt Washingtonpost.com} (ranked 9$^{th}$ in news and media publisher category in USA \cite{washingtonpost}) is one of the 383 sampled websites. It suffers a crash (a major breakage) in tracking and mixed scripts JS blocking ({\tt TMS}). On the contrary, the website is completely functional and tracking-free at method-level JS blocking ({\tt TM}). Similarly, on {\tt tenki.jp} (ranked 4$^{th}$ in the streaming and online TV category in Japan \cite{tenki}), manual inspection reveals a missing Twitter widget and a Twitter button
 in tracking and mixed scripts JS blocking ({\tt TMS}). These breakages are documented as minor breakages. However, in method-level JS blocking ({\tt TM}), all tracking advertisements are blocked, and both the button and widget appear correctly and are functional, similar to the control experiment ({\tt CTRL}). The website {\tt ndtv.com} (rank 5$^{th}$ in the news and media category in India \cite{laghari}) renders multiple advertisements in the control experiment ({\tt CTRL}). Website completely crashes in tracking and mixed scripts JS blocking ({\tt TMS}), whereas, in method-level JS blocking, it renders normally without any advertisement.

 We also argue that minor improvements can make a difference in many websites. For example, website {\tt gamestop.com} (rank 9$^{th}$ in the gaming category in USA \cite{gamestop}) shows 37.5\% breakage in tracking and mixed scripts JS blocking ({\tt TMS}) whereas shows only 12.5\% breakage at method-level JS blocking({\tt TM}). At {\tt TMS}, we see unexpected white spaces on the top of the website, a minor breakage in the appearance category. The webpage's home button also causes the website to crash, a major breakage recorded in additional functionality. However, in {\tt TM}, we only see an unexpected white space on the website, a minor breakage in the appearance category. These results  also affirm that the breakage metrics (network request and media resources) used in Phase I  are effective measures of breakage.

\subsubsection{RQ6: Is method-level blocking most effective in reducing breakage and eliminating tracking?}

Although method-level JS blocking ({\tt TM}) performs significantly better than tracking and mixed JS blocking ({\tt TMS}), there are cases where we observe little or no improvement. This is mainly because of 6\% methods still show mixed behavior \ie include tracking and functional code. {\tt Elpais.com} (currently ranks 2$^{nd}$ in the news and media publisher category in Spain \cite{eplais}) fails to load a single resource in tracking and mixed scripts JS blocking ({\tt TMS}). However, in method-level JS blocking ({\tt TM}), it causes the navigation bar to be unresponsive, a minor breakage due to the mixed method {\tt e.loadInternal} in script {\tt provider.hlsjs.js}.

 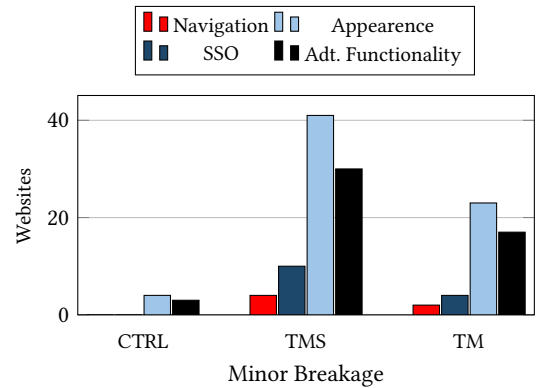
\begin{figure}[t]
        \begin{tikzpicture}
            \begin{axis}[
                width  = 0.43*\textwidth,
                height = 4.5cm,
                name=p2a,
                major x tick style = transparent,
                ybar=2*\pgflinewidth,
                ymin=0,
                bar width=10pt,
                xlabel= Minor Breakage,
                enlarge x limits=0.2,
                ymajorgrids = true,
                ylabel = {Websites},
                symbolic x coords={CTRL, TMS, TM},
                xtick = data,
                scaled y ticks = false,
                x tick label style={font=\small,text width=1.5cm,align=center},
                ylabel style={text width=1cm, font=\small ,align=center},
             legend columns=2,
                    transpose legend,
                    legend style={
                    at={(0.87,1.25)},
                    font=\small,
                    fill=none,
                    anchor= east}
            ]
                \addplot[style={fill=red,mark=none}]
                    coordinates {(CTRL,0) (TMS,4) (TM,2)};
                \addplot[style={fill=last-year,mark=none}]
                    coordinates {(CTRL,0) (TMS,10) (TM,4)};
                \addplot[style={fill=this-year,mark=none}]
                    coordinates {(CTRL,4) (TMS,41) (TM,23)};
                \addplot[style={fill=black,mark=none}]
                    coordinates {(CTRL,3) (TMS,30) (TM,17)};
                
              \legend{Navigation, SSO, Appearence, Adt. Functionality}
            \end{axis}  
        \end{tikzpicture}
        \caption{Comparison of "minor" breakage in tracking and mixed JS blocking ({\tt TMS}) vs method-level JS blocking ({\tt TM}) among 383 sampled websites. }
      \label{fig:minor}
     \vspace{-.2in}
\end{figure}   
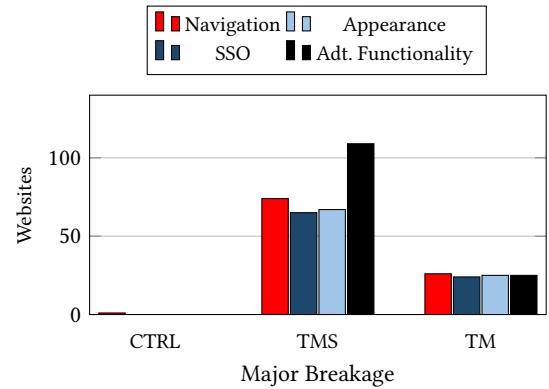
\begin{figure}[t]
        \begin{tikzpicture}
            \begin{axis}[
                width  = 0.43*\textwidth,
                height = 4.5cm,
                name=p3a,
                major x tick style = transparent,
                ybar=2*\pgflinewidth,
                bar width=10pt,
                xlabel= Major Breakage,
                enlarge x limits=0.2,
                ymajorgrids = true,
                ylabel = {Websites},
                symbolic x coords={CTRL, TMS, TM},
                xtick = data,
                ymin=0,
                ymax=140,
                scaled y ticks = false,
                x tick label style={font=\small,text width=2cm,align=center},
                ylabel style={text width=3cm, align=center, font=\small},
                legend columns=2,
                    transpose legend,
                    legend style={
                    at={(0.87,1.25)},
                    font=\small,
                    fill=none,
                    anchor= east
            }
            ]
                \addplot[style={fill=red,mark=none}]
                    coordinates {(CTRL,1) (TMS,74) (TM,26)};\addlegendentry{Navigation};
                \addplot[style={fill=last-year,mark=none}]
                    coordinates {(CTRL,0) (TMS,65) (TM,24)};\addlegendentry{SSO};
                \addplot[style={fill=this-year,mark=none}]
                    coordinates {(CTRL,0) (TMS,67) (TM,25)};\addlegendentry{Appearance};
                \addplot[style={fill=black,mark=none}]
                    coordinates {(CTRL,0) (TMS,109) (TM,25)};\addlegendentry{Adt. Functionality};
            \end{axis}
        \end{tikzpicture}
        \caption{Comparison of "major" breakage in tracking and mixed JS blocking ({\tt TMS}) vs method-level JS blocking ({\tt TM}) among 383 sampled websites.}
      \label{fig:major}
     \vspace{-4ex}
\end{figure} 
 
\subsubsection{RQ7: Can webpages sustain simply removing the tracking JS method?} 
On 100K webpages, we have found that webpages in their vanilla form have 1.32 severe errors on average. Severe error refers to three main compile-time errors in JavaScript: syntax errors, runtime errors, and logical errors. Errors are common in JS and do not always impact functionality. Compared to other software, webpages can withstand many runtime issues, such as network error, JS script not found, and JS script syntax errors that can arise from diverse host environments. In our experiments, we block JS tracking method by simply renaming the method, which may lead to MethodNotFound error.
Replacing a method name and redirecting its invocation may generate additional errors. However, such errors do not affect the website's functionality, as they only terminate the tracking-inducing thread in the JS process. 


\section{Discussion}
\label{sec: discussion}
In this section, we present the key takeaways of our empirical investigation, highlight the key challenges of effective JS blocking, and offer future ideas for dynamic analysis-based fine-grained JS blocking.

\noindent\textbf{JS blocking at finer granularity.}
While blocking JS tracking methods is beneficial, we still observe that 5.5\% webpages with some levels of tracking activity and functionality breakage. 
These webpages contain method(s) that (1) implement both tracking and functionality or (2) are used by tracking and functional code for downstream activity (\eg initiating a network request). 
We foresee better separation at a finer granularity. 
In the future, we propose applying dynamic program slicing \cite{agrawal1990dynamic,korel1988dynamic,xu2005brief} to separate tracking statements from functional statements. 
For inseparable code, we propose dynamic invariant detection \cite{ernst2007daikon,leino2004object} to construct program variable profiles for tracking and functional behaviors. 
Program in-variants for tracking can be used as an automated guard to prohibit tracking execution.

\noindent\textbf{Dynamic nature of JS.}
We find that a number of scripts use dynamic features such as {\tt eval()} and anonymous functions \cite{sarker2020hiding,moog2021statically}. 
A number of scripts also employ JS  minification and obfuscation techniques that produce code that is uninterpretable manually \cite{Skolka19minifiedobfuscatedJS,ngan2022nowhere}.  
Such practices further motivate the use of advanced dynamic program analysis techniques for tracking code identification and removal.


\noindent\textbf{JS dataflow analysis.}
In this work, we captured the stack trace of a tracking or functional network request and then annotate the script method at the top of the stack.
By focusing on request-initiator code units, we may miss opportunities to trace back to the source of the tracking behavior inside the nested JS codebase. 
Finding such a location may offer better opportunities to preserve functionality as the request-initiator method or script may simply be a ``gateway'' for all network requests. 
In addition to the call stack, we can also leverage the dataflow graph of the JS codebase to perform a richer analysis of a webpage's execution. 
%
For example, in Listing \ref{lst:mthd}, the stack trace inside the method {\tt B} does not contain the parameter {\tt C}. 
Since the method {\tt B} depends on parameter {\tt C}, the identification technique may not understand the entire context when {\tt B()} is called. 
We recommend capturing such rich execution traces with calling contexts and a complete data flow graph to understand better the flow of information through the nested code and how it influences the execution behavior, tracking, or functional. 
We anticipate that such traces can help identify better locations (\eg non request-initiator methods) to alleviate tracking while preserving functionality.

\begin{lstlisting}[caption={Call stack does not show complete dataflow.},label={lst:mthd}]
function TrackingReq() {
    C = getVal();
    B(C)}; };
\end{lstlisting}

\noindent\textbf{Performance impact of JS blocking.}
Although we do not consider performance in our analysis, our focus is to minimize tracking without comprising functionality.
Recent works \cite{chaqfeh2022jsanalyzer, chaqfeh2020jscleaner} show that the removal of non-critical components of JS code can significantly reduce page load times. 
Similarly, removing the tracking JS code may reduce the performance overhead along with functionality preservation.

\noindent\textbf{Other future research directions.} 
We plan to conduct an investigation into more meaningful and semantics-aware tracking code identification. 
Our key observation is that finding a tracking code unit in webpages has striking similarities with fault localization.
Even a simple faulty code localization method such as SBFL showed promising results towards functionality-preserving JS blocking. 
On the code refactoring front, our observation of 100K vanilla live websites reveals that today's webpages can withstand severe errors. 
Therefore, we expect that slightly unsafe code refactoring techniques to remove the tracking code may be promising in effectively preserving functionality while preventing tracking.

Future tracking code identification techniques can greatly benefit from recent advances in automated debugging and fault localization~\cite{masri2015automated, janssen2009zoltar}. 
For example, given filter list as a test oracle, we can adapt search-based debugging approaches to perform a systematic search on JS code and precisely isolate the tracking and functional code units \cite{misherghi2006hdd}.
Similarly, the completeness of static code dependency analysis (\eg reachability analysis) can complement the soundness of dynamic analysis (\eg call graph) to improve the precision of tracking code localization.

Code clone detection is an active area of research, with many advanced techniques available for traditional software ~\cite{codeclones}. 
Given annotated JS code units, code clone detection techniques can identify similar code on webpages to find the presence of tracking code. 
Once a JS code clone is correctly detected, we can leverage supervised learning  ~\cite{webgraph,iqbal21fpinspector} to extract valuable features, both semantic and syntactic, for accurate tracking code localization. 
If such an accurate model is available, a JS blocker can detect tracking JS code units in real time and block them before loading the website.

Similar to training a classification model, one possible direction is to create a taxonomy of tracking code's signature, similar to the ones in malware detection~\cite{jang2011bitshred, hu2013mutantx, zhang2020dynamic}, and find a match with a webpage's JS entity at page load. 
However, page load times are critical in the web domain, refraining from any computationally expensive operation. 
Using fingerprints to locate tracking code at page load is a lightweight process that can easily be performed at page load time without a noticeable slowdown.


\paragraph{Can publishers also benefit from the results of our JS blocking study?}  
Our study is conducted from the perspective of privacy-enhancing content-blocking tools. 
%
%
%
If suitable, we suggest publishers adopt an approach such that either the website works reasonably without JS or at least employ a highly decoupled JS architecture that separates tracking and functionality, \ie separate JS scripts/methods. 
This architecture will retain functionality effectively when JS code level blocking reduces tracking. 
On the contrary, publishers who want to retain maximum tracking may leverage the current weakness of JS script-level content blocking by maximizing the overlap between tracking and functional code units.

\section{Limitations}
\label{sec: limitations}

\noindent\textbf{Internal validity.}
Our analysis in Section~\ref{sec: results} relies on the correlation between a JS blocking strategy and the webpage's behavior in terms of network requests and resource loads. 
However, other confounding factors may impact the webpage's behavior. 
For instance, some webpages fetch different number and type of resources for each visit due to the inherent dynamism. 
For our experiments, requests monitored in one experiment may not be triggered in another experiment. 
Other factors include behavior change due to environment (\ie browser and host OS), visit time, and location. 
We minimize internal validity threats by keeping the environments consistent across different blocking configurations \ie same location, browser, and stateless crawls.

\noindent\textbf{External validity.} 
We conducted our experiments using the Chro\-me browser with a Chrome-based extension. 
Extensions on other browsers have different permissions and have access to a varying set of information about a webpage's behavior. 
While our choice of using the Chrome browser minimizes external validity threats, it is possible that our results may not fully generalize to JS method-level blocking on other browsers. 
Similarly, our annotation relies on previously observed tracking behavior captured in filter lists.
Its effectiveness may be limited for unseen JS. 
To minimize this issue, we use the two most actively maintained filter lists for annotation.

\noindent\textbf{Construct validity.}
We collect the website's data at page load time and do not capture other events triggered by the user interactions such as scrolling and clicking. This is a general limitation of dynamic analysis that can be mitigated by using a forced execution framework \cite{kim2017j}.
%

\section{Related Work}
\label{sec: related work}


Smith et al. \cite{sugarcoat} and Amjad et al. \cite{trackersift} identify tracking code regions in the JS scripts of websites. Sugarcoat \cite{sugarcoat} dynamically captures the call graphs of web APIs and uses it to determine the call site in JS code that tries to access the user-sensitive information from the local storage, which is unanimously considered as tracking behavior. They replace these identified call sites with the surrogate JS code that mitigates the information access but preserves functionality. This process helps create surrogate scripts for exception rules in the filter list. SugarCoat requires excessive manual effort by a domain expert to identify the tracking call sites in the JS code. Due to this limitation, our empirical study could not validate SugarCoat's effectiveness. 
Amjad et al. \cite{trackersift} introduce a hierarchical approach to annotate web entities (domain, hostname, script, and method) to precisely isolate the code responsible for tracking behavior. They dynamically collect the call stack information for tracking behavior and isolate the entities based on their participation in invoking it. We adapt their approach for web corpus collection and extend it to enable real-time code blocking and capture additional information. 

Modern websites extensively use third-party JS scripts that may access potentially sensitive information \cite{js-1, jueckstock2019visiblev8, js-tracker, li2022towards}. Tran et al. \cite{js-tracker} develop a principal-based tainting approach that dynamically analyzes the JS libraries to identify the underlying privacy violations. They tag each compiled JS library at run-time and observe its suspicious behavior with the author-defined principles \ie a set of permissions that should not be violated. Similarly, Staicu et al. \cite{js-1} introduce an automated approach that collects taint specifications of JS libraries and identifies behaviors that lead to security vulnerabilities. These approaches work at the granularity of JS libraries, which, as we find, is insufficient for preserving functionality.
Moreover, these works use taint analysis that incurs prohibitively high-performance overhead and can not efficiently work in the browser in real-time. Prior work's findings on the challenges from JS dynamism resonate with our findings. Jueckstock et al. \cite{jueckstock2019visiblev8} present a lightweight dynamic analysis tool using chrome V8 to identify untrustworthy JS scripts. It logs function calls and storage access during JS execution to identify suspicious code.

The limitations of identifying tracking code share similarities with prior research on fault localization.
For example, spectra-based fault localization (SBFL) \cite{sb2,pearson,Souza, sarhan2022survey,widyasari2022xai4fl} leverages the statement coverage using the set of passing and failing test cases to localize the statement that is most likely to induce a test failure.
Similarly, Bela et al. \cite{vancsics2021call} and Laghari et al. \cite{laghari} present an approach that uses the frequency of method occurrence in the call stack of failing test cases for localizing the faulty methods.
A method that appears more in the call stack of failing test cases is more likely to be faulty.
Abreu et al. \cite{Zoe} conducted an empirical study on the accuracy of these SBFL techniques and highlighted that these approaches are independent of the quality of the test oracle.
Crowdsourced blocklists~\cite{EasyList1, EasyPrivacy, fanboy, cookies} are the authoritative source of labels for requests and are adequate to detect tracking behavior.
%
%

%
Websites heavily rely on JS libraries containing significant dead code that is unused or unreachable, posing a noticeable impact on the website's performance.
Kupoluyi et al.~\cite{kupoluyi2021muzeel} highlight that popular websites have 70\% unused functions, and their elimination can speed up the page load by 30\%.
Recent works \cite{chaqfeh2022jsanalyzer, chaqfeh2020jscleaner} have further explored non-critical regions in JS libraries and the performance overhead caused by them.
Zaki et al. \cite{chaqfeh2020jscleaner} propose on rule-based classification techniques to identify and replace non-critical regions in JS using pre-define code patterns,  achieving a 50\% reduction in page load time.
Towards the same goal, Chaqfeh et al. ~\cite{chaqfeh2022jsanalyzer} develop a tool that helps developers in eliminating JS elements by visually inspecting them and shows 90\% improvement in Google's lighthouse performance score.
Similarly, Vazquez et al. \cite{vazquez2019slimming} proposed a technique to decompose bundles JS code in a website, reducing code size by 26\%.
Our findings in this study are equally beneficial to the research on improving website performance and energy consumption that specifically adopt functionality-preserving code debloating approaches.

\section{Conclusion}
\label{sec: conclusion}
In this paper, we conduct a large-scale empirical investigation on the impact of different JS Code blocking methodologies on 100K websites, followed by a careful visual inspection of 383 websites to measure website breakage. Our results show that blanket JS blocking prevents tracking but incurs major functionality breakage on approximately two-thirds of the websites. We identify that 15\% of the scripts on the web combine tracking and functionality, leading to website breakage if blocked. When we increase the granularity of JS blocking to target tracking methods inside mixed scripts, the functional breakage of websites reduces by 2X while providing the same level of tracking prevention. Our in-depth manual inspection of 383 websites validates that method-level JS blocking reduces major breakage by 3.8$\times$. Through this study, we highlight the promise of fine-grained JS blocking and the subsequent open challenges towards adapting such a technique in practice.

\begin{acks}
This work is partly supported by the National Science Foundation under grant numbers 2106420, 2103038, 2138139, 2103439, and 2051592.
We want to thank the anonymous shepherd and reviewers for their constructive feedback that helped improve the work. 
\end{acks}

\bibliographystyle{ACM-Reference-Format}
\bibliography{sample-base}










\end{document}